\documentclass[pra,superscriptaddress,aps,showpacs,floatfix,nofootinbib,twocolumn,reprint]{revtex4-1}
\usepackage{epsfig}
\usepackage{amssymb}
\usepackage{amsmath}
\usepackage{amsfonts}
\usepackage{xcolor,graphicx}

\usepackage{hyperref} 
\newcommand{\unit}{1\!\!1}

\usepackage{comment} 

\newcommand{\be}{\begin{equation}}
\newcommand{\ee}{\end{equation}}

\definecolor{violet}{RGB}{50,40,200}

\newcommand{\ssec}[1]{\emph{#1}.---}

\definecolor{DarkBlue}{RGB}{10,10,140}

\begin{document}

\title{Pseudogap regime of the unitary Fermi gas with lattice auxiliary-field quantum Monte Carlo in the continuum limit}

\author{S.~Jensen}
\affiliation{Center for Theoretical Physics, Sloane Physics Laboratory, Yale University, New Haven, Connecticut 06520, USA}
\affiliation{Department of Physics, University of Illinois at Urbana-Champaign, Urbana, Illinois 61801, USA}
\author{C. N.~Gilbreth}
\affiliation{Quantinuum, Broomfield, Colorado 80021, USA}
\author{Y. Alhassid}
\affiliation{Center for Theoretical Physics, Sloane Physics Laboratory, Yale University, New Haven, Connecticut 06520, USA}

\begin{abstract}
The unitary Fermi gas (UFG) is a strongly correlated system of two-species (spin-1/2) fermions with a short-range attractive interaction modeled by a contact interaction and has attracted much interest across different disciplines. The UFG  is considered a paradigm for strongly correlated superfluids and has been investigated extensively, with generally good agreement found between theory and experiment.  However, the extent of a pseudogap regime above the critical temperature $T_c$ for superfluidity, in which pairing correlations persist, is still debated both theoretically and experimentally. Here we study thermodynamic properties of the UFG across the superfluid phase transition using finite-temperature lattice auxiliary-field quantum Monte Carlo (AFMC) methods in the canonical ensemble of fixed particle numbers. We present results for the condensate fraction, a model-independent energy-staggering pairing gap, the spin susceptibility,  and a free energy-staggering pairing gap.  We extrapolate our lattice AFMC results to the continuous time and continuum limits, thus removing the systematic error associated with the finite filling factor of previous AFMC studies. Applying a finite-size scaling analysis to the condensate fraction results, we determine the critical temperature to be $T_c=0.16(1)\, T_{F}$.  For the largest particle number studied $N=114$, the energy-staggering pairing gap, which provides a direct signature of a pseudogap regime, is suppressed above a pairing scale temperature of $T^{*}\approx 0.2\,T_F$. The spin susceptibility displays moderate suppression above $T_c$ with a spin gap temperature of $T_s\approx 0.2 \,T_F$. We also calculate a free energy-staggering pairing gap, which shows substantially reduced statistical errors when  compared  with the energy-staggering gap, allowing for a clear signature of pairing correlations in the finite-size system. All results indicate that the pseudogap regime is narrow, with pseudogap signatures emerging at temperatures below $T^{*}\approx 0.2 \, T_F$. The reduced statistical errors of the free energy gap enable an extrapolation at low temperatures, allowing an estimate of the zero-temperature pairing gap $\Delta_E = 0.576(24) \, \epsilon_F$. 
\end{abstract}

\maketitle

\section{Introduction}
The unitary Fermi gas (UFG), with diverging s-wave scattering length, is of great interest in diverse areas of physics as it provides a well-defined paradigm for strongly correlated systems relevant to the study of nuclei, quark matter, neutron stars, and atomic Fermi gases. It undergoes a $U(1)$ phase transition from a normal strongly correlated gas to an s-wave superfluid~\cite{Leggett2006, Bloch2008,Giorgini2008, Inguscio2008,Zwerger2012, Levin2012, Svistunov2015,Pitaevskii2016} below a critical temperature $T_c$. The UFG is realized in the middle of a crossover from a weakly interacting Fermi gas described by the Bardeen-Cooper-Schrieffer (BCS)~\cite{Bardeen1957} theory to a weakly repulsive Bose-Einstein condensate (BEC)~\cite{Zwerger2012, Eagles1969, Leggett1980, Nozieres1985} (for $T<T_c$ in the superfluid phase). The UFG has remarkable properties as one of the highest-temperature superfluids in units of the Fermi temperature $T_{F}$~\cite{Randeria2010}. It exhibits a nearly perfect fluid behavior~\cite{Enss2012-2, Schafer2017, Nishida2005}, and can be realized  in ultracold atomic Fermi gas experiments with $^{6}\textrm{Li}$ and $^{40}\textrm{K}$~\cite{Ketterle2008, Hu2022}.

There is no known controlled expansion in a small parameter (around some nearby exact solution) to approach the superfluid transition of the UFG. It is therefore difficult to theoretically address this system on a quantitative level. The UFG, due in part to its simplicity and broad interest, has become a testing ground for strongly correlated many-body methods. Successful approaches in advancing our understanding of the UFG include the epsilon expansion~\cite{Nishida2006}, the Luttinger-Ward approach~\cite{Haussmann2007,Haussmann2009,Zwerger2016,Frank2018}, conformal field theory methods~\cite{Nishida2007-1, Nishida2007-2}, quantum critical approaches~\cite{Sachdev2007, Enss2012-2}, non-self consistent diagrammatic resummation methods~\cite{Perali2002,Ohashi2002, Chen2005}, and various quantum Monte Carlo methods~\cite{Gilbreth2013, Bulgac2006, Bulgac2008, Drut2012, Akkineni2007,Burovski2006, Burovski2006-2, Burovski2008, Goulko2010, Van2011, Hu2010, Lee2006, Endres2013}. 

The UFG has been extensively investigated experimentally in recent decades using ultracold Fermi gases ~\cite{Regal2004, Zwierlein2004, Kinast2004, Chin2004, Zwierlein2005, Navon2009, Navon2011, Ku2012, Zwerger2012, Zwierlein2017}, but important open problems remain. A topic that has generated considerable interest is the existence and extent of a pseudogap regime above the critical temperature $T_c$ for superfluidity, in which there is no off-diagonal-long-range order (ODLRO) but pairing correlations persist with a gapped single-particle spectrum near the Fermi surface. 

The extent of pairing correlations in the UFG has been debated in recent years. For low-temperatures ($T \to 0$), studies of pairing properties demonstrated good  agreement between theory and experiment~\cite{Shin2007, Gezerlis2008, Carlson2003, Chang2004, Carlson2005, Carlson2008, Carlson2011}.  However, at finite temperature, and in particular across the superfluid critical temperature $T_c$, our understanding is more limited. While there has been substantial progress in our understanding of the normal phase both theoretically~\cite{Ohashi2012, Kashimura2012, van_Wyk2016, Strinati2002, Strinati2011, Strinati2012-2, Tajima2016, Pini2019, Perali2002, Perali2011, Palestini2012, Pantel2014, Tajima2014,Santos1994, Janko1997, Sewer2002, Chien2010a,Chien2010b,Wlazlowski2013,Magierski2009,Magierski2011,Enss2012,Jensen2020a} and experimentally~\cite{Gaebler2010,Ku2012,Navon2009,Navon2011, Sommer2011, Wulin2011}, our understanding of pairing correlations  (even qualitatively) above $T_c$ is still incomplete. For reviews of the pseudogap regime in the UFG, see Refs.~\cite{Chen2005, Chen2014, Mueller2017, Jensen2019}. 

In this work, we calculate several thermodynamic properties of the spin-balanced homogeneous UFG using lattice auxiliary-field quantum Monte Carlo (AFMC) in the canonical ensemble of fixed particle numbers.  In particular, we calculate the condensate fraction, a model-independent energy-staggering pairing gap,  the spin susceptibility, and the free energy-staggering pairing gap for $N=40,66$ and $114$ particles. All properties are calculated in the continuum limit in both imaginary time and lattice spacing, thus removing systematic errors of a finite-filling factor of our previous work~\cite{Jensen2020a} and allowing direct comparison with experiments.  

Applying finite-size scaling to the condensate fraction data, we determine the superfluid critical temperature of the UFG to be $T_c=0.16(1)T_{F}$. The canonical ensemble framework enables the calculation of an energy-staggering pairing gap and a free energy-staggering pairing gap without the need for a numerically challenging analytic continuation. In particular, the free energy-staggering gap has remarkably small statistical errors, allowing for an extrapolation from finite temperature to estimate accurately the $T=0$ energy-staggering pairing gap $\Delta_E/\varepsilon_{F} = 0.576(24)$  (where $\varepsilon_{F}$ is the Fermi energy). From the pairing gaps for our largest particle number simulated, $N=114$, we estimate an upper bound on the pairing temperature scale associated with these observables to be $T^{*} \approx 0.2 T_F$ (this pairing temperature decreases with increasing particle number $N$). We find a similar value for the spin gap temperature of the spin susceptibility (for $N=114$), $T_s \approx 0.2 T_F$. These results for $T^*$ and $T_s$  indicate a narrow temperature pseudogap regime for the UFG.

The outline of this paper is as follows. In Sec.~\ref{UFG} we provide background material for the UFG and discuss progress toward understanding the pseudogap regime, with a focus on recent lattice Monte Carlo studies. In Sec.~\ref{sec:AFMC} we describe our canonical-ensemble AFMC method, and the recent optimization algorithms that allow for large lattice simulations. In Sec.~\ref{sec:Results} we present the AFMC results of this work, and  in Sec.~\ref{sec:conclusions} we provide concluding remarks. 

\section{BCS-BEC crossover and the UFG}\label{UFG}

The BCS-BEC crossover in three spatial dimensions is a continuous crossover (at temperature $T=0$) as a function of  $(k_{F}a)^{-1}$ (where $a$ is the s-wave scattering length and $k_{F}$ is the Fermi wavenumber) from a weakly attractive Fermi gas described by BCS theory to a system of weakly interacting bosons described by BEC theory~\cite{Zwerger2012, Giorgini2008}. As temperature is increased, the gas exhibits a $U(1)$ (3D XY) superfluid phase transition into a normal disordered phase. The BCS regime corresponds to $(k_F a)^{-1} \sim - \infty$ where the normal phase can be described by Fermi liquid theory. The BEC regime, corresponding to  $(k_F a)^{-1} \sim +\infty$, is a weakly interacting Bose gas of tightly bound dimers with binding energy $\epsilon_{b}= \hbar^2/(m a^2)$ where $m$ is the mass of a single fermion. In the BCS regime, for $T<T_c$ there are Cooper pairs of spatial extent much greater than the interparticle separation $\rho^{-1/3}$ where $\rho=N/V$ is the particle density for $N$ fermions in a spatial volume $V$. In the BEC regime, the pairs are tightly bound dimers of size much smaller than the interparticle separation. In the unitary limit, defined by $(k_F a)^{-1} = 0$ (i.e., infinite scattering length), the two-particle bound state is a zero-energy resonance. In this regime, the size of the Cooper pairs is comparable to the interparticle separation and the system is strongly correlated.

\subsection{Scattering amplitude and the continuum model}\label{sec:define_problem}

We consider a dilute Fermi gas at low temperature. The scattering between two particles with wavevectors $\bold{k}_{1}$ and $\bold{k}_{2}$ is well described by the low-momentum s-wave expansion of the inverse of the scattering amplitude $f(k)$,
\begin{equation}\label{UFG-amplitude}
f^{-1}(k)=-1/a -ik +r_{e}k^{2}/2 + \ldots \;,
\end{equation}
where $\bold{k}$ is the relative momentum wavevector $\bold{k}=\bold{k}_{1}-\bold{k}_{2}$, $r_{e}$ the effective range, and $\dots$ refers to higher-order corrections in $k^2$.  This expansion holds for $^{6}$Li and $^{40}$K, where the low-energy interaction at large distance is well described with the van der Waals potential $V(r)=-C_{6}/r^6$ with a microscopic interparticle interaction range of $R_0=(mC_{6}/\hbar^2)^{1/4}$  ($\hbar$ is Planck's constant) in the limits $R_0<< 1/k_{F}, \lambda_{T}$. Here $\lambda_T$ is the thermal de Broglie wavelength $\lambda_{T}=\sqrt{2\pi\hbar^{2}/mk_{B}T}$ ($k_B$ is the Boltzmann constant), and the Fermi wavenumber is given by $k_{F}=(3\pi^{2}\rho)^{1/3}$. See Ref.~\cite{Khuri2009} for a detailed discussion of the low-momentum scattering expansion. Though the effective range parameter $r_{e}$ is often of the order $R_0$ (for a broad Feshbach resonance $|r_{e}|\sim R_{o}$)  this is not guaranteed to hold~\cite{Castin2012, Werner2012}. The unitary limit is reached when the modulus of the scattering amplitude is maximized, saturating the constraint imposed by unitarity, from the optical theorem, $\textrm{Im} f(k) = k |f(k)|^{2}$, and leading to $f(k)=-1/ik$. We thus define the UFG by the conditions: $R_0<<1/k_{F}$, $k_{F}a\rightarrow 0$, $r_{e}k_{F}<<1$, and $R_0<<\lambda_{T}$.

These conditions are met with cold atomic Fermi gases using Feshbach resonance techniques to tune an external magnetic field so that a closed channel bound state becomes resonant with open channel scattering (see Refs.~\cite{Chin2004, Chin2010, Giorgini2008}). Near a broad Feshbach resonance, many properties of the gas can be understood with a single-channel model using a contact interaction between spin-up and spin-down fermions (corresponding to an atomic hyperfine state manifold with two relevant states). More specifically, we model a system of $N_{\uparrow}$ spin-up, and $N_{\downarrow}$ spin-down, fermions in a three-dimensional cubic box of side length $L$ with periodic boundary conditions.  The continuum Hamiltonian of this system is given by
\begin{multline}\label{Hamiltonian}
\hat{H} = \sum_{s_z} \int d^3 r \hat{\psi}^{\dagger}_{s_z}(\bold{r})\left(-\frac{\hbar^{2} \nabla^{2}}{2 m}\right)\hat{\psi}_{s_z}(\bold{r}) \\
+V_0 \int d^3 r \hat{\psi}^{\dagger}_{\uparrow}(\bold{r})\hat{\psi}^{\dagger}_{\downarrow}(\bold{r})\hat{\psi}_{\downarrow}(\bold{r})\hat{\psi}_{\uparrow}(\bold{r}) \;,
\end{multline}
where  $\hat{\psi}^{\dagger}_{s_z}(\bold{r})$ and  $\hat{\psi}_{s_z}(\bold{r})$ are, respectively, creation and annihilation operators for a fermion at position $\bold{r}$ with spin projection $s_z$. The contact interaction model is ill-defined without regularization. The model can be regularized with an ultra-violet (UV) cutoff $\Lambda$ in momentum space. 

\subsection{The pseudogap regime}

A pseudogap regime is known to exist in high-$T_c$ superconductors. For instance, the specific heat is strongly suppressed above $T_c$ in contrast to conventional superconductors which show Fermi liquid behavior for low temperatures above $T_c$~\cite{Dagotto1994,Timusk1999}. The mechanism leading to the pseudogap regime of the cuprates is still not fully understood~\cite{Randeria2010,Dagotto1994,Timusk1999} as the complexity, competing orders, and strongly correlated nature of the problem are difficult to address with theoretical approaches. The BCS-BEC crossover is also strongly correlated for $|k_{F}a|>1$ but the ordering is well understood to be a $U(1)$ superfluid phase transition with no competing orderings and the appropriate microscopic model is simple. Pseudogap phenomena in this system can be attributed to precursor pairing correlations.  One can define a temperature scale $T^*$ as the temperature at which signatures of pairing first appear as the temperature is lowered from the normal phase. There are different definitions of $T^*$ that depend on the choice of observable used to determine the signatures of pairing correlations~\cite{Mueller2017, Jensen2019}. 
Precursor pairing develops in the BCS-BEC crossover for $T>T_c$ as the coupling changes from the BCS regime where $T^{*}=T_c$~\cite{Fetter1971} to the BEC regime where $T^{*}>>T_{c}$ and is the dissociation temperature scale for breaking the two-body bound dimers due to thermal fluctuations. However, it is not clear how this precursor pairing emerges. It is a challenging many-body problem to calculate the extent of pairing signatures above $T_c$ for $|k_{F}a|>1$. 

Many different techniques have been applied to the UFG and the calculation of observables that are expected to show pairing structure. The AFMC studies of Refs.~\cite{Magierski2009, Magierski2011, Wlazlowski2013}, where a spherical cutoff was used in the lattice model, found signatures of pairing above the critical temperature. Refs.~\cite{Chien2010a, Chien2010b, Perali2011, Kashimura2012, Palestini2012-2, Palestini2012, Tajima2014, Wulin2011, Pini2019, Pini2020} applied non-self-consistent diagrammatic approaches and found varying degrees of pairing above the critical temperature and gapped signatures in support of a pronounced pseudogap regime for the UFG.  In contrast to these works, the self-consistent Luttinger-Ward diagrammatic approach of Refs.~\cite{Haussmann2009, Enss2012} did not find a pronounced pseudogap regime and found that $T^{*}$ and $T_c$ are very close for the UFG (see Refs.~\cite{Pini2019, Pini2020} for comparisons between self-consistent and non-self-consistent approaches). On the experimental side, the results also show disagreement. The works of Refs.~\cite{Wulin2011, Gaebler2010} found pseudogap signatures whereas Refs.~\cite{Ku2012, Navon2009, Navon2011, Sommer2011} saw no gapped signatures.

In the UFG, the value of $T^*$ ($T^* \geq T_c$) is still debated in the literature.  Discrepancies exist even when the same pairing observables are used to determine $T^*$,  as the temperature dependence of these observables differ substantially between various works. In Sec.~\ref{sec:Results} we present controlled AFMC results for two pseudogap signatures: the temperature dependence of the spin susceptibility and the energy-staggering pairing gap across the superfluid phase transition. 

\subsection{Lattice Model}\label{sec:lattice_model}

The continuum model  described by the Hamiltonian in Eq.~(\ref{Hamiltonian}) can be replaced by a discrete lattice model and the continuum limit is then recovered by extrapolating to zero filling factor.  

To construct the lattice model, we discretize a cubic box of volume $V=L^3$ with $N_{L}^{3}$ lattice sites forming a three-dimensional cubic lattice with lattice spacing $\delta x = L/N_L$. The lattice of linear size $N_L \in \mathbb{Z}^{+}$ has lattice sites at positions $x=0,\pm\delta x, \pm 2\delta x,...,\pm M\delta x$ when $N_{L} = 2M + 1$ is odd.  The lattice Hamiltonian for a contact on-site attractive interaction is given by
\begin{equation}\label{Hamiltonian_Lat}
\hat{H}=\sum_{{\bf k},s_z }\epsilon _{\bf{k}}\hat{a}^{\dagger }_{{\bf k},s_z }\hat{a}_{{\bf k},s_z}+g\sum_{\bf{x}}\hat{n}_{\bf{x},\uparrow}\hat{n}_{\bf{x},\downarrow} \;,
\end{equation}
where $\hat{a}^{\dagger }_{{\bf k},s_z}$ and $\hat{a}_{{\bf k},s_z}$ are, respectively,  creation and annihilation operators for discrete quasi-momentum $\hbar\bold{k}$ and spin projection $s_z$. In this work we use the quadratic single-particle dispersion relation $\epsilon_{\bold{k}}=\hbar^{2}k^2/2m$ though other dispersion relations have also been used (see Ref.~\cite{Werner2012} for a discussion of dispersion relations). The lattice operators obey the anticommutation relations $ \{  \hat{a}_{{\bf k},s_z},\hat{a}^{\dagger}_{{\bf k}',s_z'}\}=\delta_{{\bf k},{\bf k}'}\delta_{s_z, s_z'}$. The number operator at lattice site $\bold{x}$ and spin projection $s_z$ is $\hat{n}_{{\bf x},s_z}=\hat{\psi}^{\dagger}_{{\bf x},s_z}\hat{\psi}_{{\bf x},s_z}$ with the lattice field operators obeying $\{\hat{\psi}_{{\bf x}, s_z}, \hat{\psi}^{\dagger}_{{\bf x}',s_z'}\}= \delta_{\bf{x},\bf{x}'}\delta_{s_z,s_z'}$.  The constant $g=V_0/(\delta x)^{3}$ in Eq.(\ref{Hamiltonian_Lat}) is the rescaled coupling constant. The bare coupling constant $V_0$ is determined by the requirement that the physical $s$-wave scattering length $a$  ($a\rightarrow \infty$ for the UFG) is reproduced on the lattice. This leads to
\begin{equation} \label{geqn}
\frac{1}{V_0}=\frac{m}{4\pi \hbar^2 a}-\int_{B}\frac{d^{3}k}{(2\pi)^{3}2\epsilon_{\mathbf{k}}} \;,
\end{equation}
where the integral in (\ref{geqn}) is over the first Brillouin zone $B$ of the lattice. 
Relation (\ref{geqn}) can be derived by solving the Lippmann-Schwinger equation on an infinite lattice. 

The UFG is recovered by solving the lattice model for $a\rightarrow \infty$  and then taking the continuum limit when the filling factor on the lattice $\nu=N/N_{L}^3\rightarrow 0$ for given particle number $N$, followed by the thermodynamic limit $N \rightarrow \infty$. The approach to the continuum limit was first addressed in the works of Refs.~\cite{Burovski2006-2,Pricoupenko2007}.  Following these works, we expect the many-body energies to be linear in $\nu^{1/3}$ for a low filling factor $\nu$. In the present work (see Sec.~\ref{sec:Results}), we take the continuum limit $\nu \rightarrow 0$ for various thermodynamic observables using a linear fit in $\nu^{1/3}$ for a given particle number $N$.  

\subsection{Lattice AFMC studies}\label{sec:AFMC_pseudogap}

Lattice AFMC methods are among the most accurate, but the main challenges have been (i) taking the continuum limit $\delta x \to 0$ (or equivalently $\nu \to 0$), and (ii) taking the thermodynamic limit of large particle number $N$. There have been several works using large-scale lattice AFMC methods to study signatures of pairing  correlations in the UFG at finite temperature. These studies have improved the understanding of finite-temperature properties of the UFG. However, satisfactory control of systematic errors has not been achieved in these studies. We summarize the current status and recent works in this section.\\

\ssec{Spherical cutoff studies}
In Refs.~\cite{Bulgac2006, Bulgac2008, Burovski2008, Magierski2009, Magierski2011, Wlazlowski2012, Wlazlowski2013, Wlazlowski2013-2, Wlazlowski2015} the UFG was studied using a spherical cutoff in the single-particle quasi-momentum space. Ref.~\cite{Burovski2008} studied a continuum model where the single-particle quasi-momentum states in the lab frame were restricted to have magnitude less than a cutoff $\hbar\Lambda$ with a renormalization condition similar to Eq.~(\ref{geqn}) but where the integration region is determined by a spherical cutoff in quasi-momentum space (i.e., $\epsilon_\bold{k}=\hbar^2 k^2/(2m)$ for $|\bold{k}|<\Lambda$ and $\epsilon_\bold{k}=\infty$ for $|\bold{k}|>\Lambda$). This model does not allow scattering to lab frame quasi-momenta states with $|\bold{k}| > \Lambda$. Refs.~\cite{Bulgac2006, Bulgac2008, Magierski2009, Magierski2011, Wlazlowski2012, Wlazlowski2013, Wlazlowski2013-2, Wlazlowski2015} used a similar model but on a periodic lattice. Pseudogap signatures at unitarity were observed with this model in the simulations of Refs.~\cite{Magierski2009, Magierski2011, Wlazlowski2013} for temperatures below $\approx 0.25\;T_F$. 

Ref.~\cite{Werner2012} calculated the two-particle scattering in the continuum limit using a spherical cutoff model and found that the spherical cutoff does not produce the desired two-body scattering amplitude for finite center-of-mass momentum $\bold{K}$. The low-energy inverse scattering amplitude was found to have the form $f^{-1}(k,K)= -1/a-ik+K/(2\pi)+\ldots$ where $K=|\bold{K}|$ (typical $k$ and $K$ are of order $k_{F}$ for scattering of the degenerate UFG).  Comparing this with Eq.~(\ref{UFG-amplitude}), the extra center-of-mass term, $K/(2\pi)$, shifts the scattering length to an effective value, $1/a_{\textrm{eff}}=1/a-K /(2\pi)$ away from unitarity ($1/a=0$) towards the BCS side. This argument suggests the spherical cutoff model underestimates pseudogap effects. However, collision effects for varying center-of-mass momentum in the spherical cutoff model lead to enhanced pairing effects~\cite{Jensen2019, Jensen2020a}. 

Ref.~\cite{Jensen2020a} performed a many-body calculation with and without the spherical cutoff. The comparison presented there is not sufficient to address differences at the many-body level in the continuum limit. The effective range parameter $r_{e}$ and the center-of-mass parameter $R_{e}$ are different when using a spherical cutoff, as opposed to including the full first Brillouin zone (i.e., they approach the continuum limit differently). To address this, simulations were performed in Ref.~\cite{Jensen2019} for the spin susceptibility for $N=20$ particles with multiple lattice sizes to approach the continuum limit. These studies, however, do not resolve the magnitude of the discrepancy in the thermodynamic limit, which has not been addressed. 

The spherical cutoff model reduces the single-particle model space and allows for larger lattice simulations within the AFMC framework to more closely approach the continuum and thermodynamic limits (for given computational resources). The early works utilizing this truncation provided the first lattice AFMC investigation of finite temperature pairing signatures. However, recent algorithm advances~\cite{Drut2011, Jensen2019, Richie2020, Gilbreth2021} and increased availability of computational resources have allowed for large-lattice simulations using the full first Brillouin zone for the single-particle model space.\\

\ssec{Full first Brillouin zone studies}
There have been a number of studies with finite-temperature AFMC using the full first Brillouin zone with the quadratic single-particle dispersion relation. The early work of Ref.~\cite{Lee2006} performed finite filling factor calculations to find a number of observables including the thermal energy and spin susceptibility. This study was performed for smaller lattice sizes up to $N_L=6$ and down to temperatures of $T/T_F=0.14$ providing the first lattice AFMC estimate of the Bertsch parameter $0.07 \le \xi \le 0.44$. Ref.~\cite{Drut2011} produced the first large-scale ($N_L > 13$) finite-temperature AFMC calculations for the UFG using the full first Brillouin zone (see Ref.~\cite{Drut2013} for details on algorithm advances).  However, this work presented results for the contact and did not focus on pseudogap properties. 
 
Recently, Ref.~\cite{Richie2020} calculated pairing observables for $N\approx 30$ for multiple values of the interaction parameter $1/(k_F a_s)$ along the BCS-BEC crossover and obtained estimates for $T^*$ as a function of $1/(k_F a_s)$. This work reached much smaller filling factors compared with the simulations of Ref.~\cite{Jensen2020a} and provided estimates throughout the entire strongly correlated regime of the BCS-BEC crossover. Ref.~\cite{Richie2020} restricted the study to a moderate particle number $N\approx 30$ and did not take the lattice continuum limit.

In Refs.~\cite{Jensen2020a, Jensen2019}, the current authors used the canonical ensemble algorithm discussed in Sec.~\ref{sec:AFMC}, to produce a large-scale study at finite filling factor $\nu \approx 0.6$ for pairing observables using the full first Brillouin zone. These works presented results for the spin susceptibility, condensate fraction, and the energy-staggering pairing gap. Results were presented for multiple particle numbers up to $N=130$. This work found a negligible pseudogap regime based on the analysis of the pairing gap and spin susceptibility for large particle number. However, these results had sizable lattice effects with an effective range parameter $k_F r_e \simeq 0.41$ and cannot be directly compared with experiment or other UFG results. This presents a large systematic error relative to the UFG with vanishing effective range parameters as defined in Sec.~\ref{sec:define_problem}. The current work (see results in Sec.~\ref{sec:Results}) addresses the systematic errors of these previous studies.

\section{Finite-temperature lattice AFMC}\label{sec:AFMC}
We calculate thermodynamic properties of the system with the Hamiltonian of Eq.~(\ref{Hamiltonian}) using a discrete lattice of size $N_{L}^3$ and total particle number $N=N_{\uparrow}+N_{\downarrow}$ with controllable systematic errors as discussed in Sec.~\ref{sec:lattice_model}. In particular, we have implemented finite-temperature auxiliary-field quantum Monte Carlo (AFMC) methods in the canonical ensemble of fixed particle numbers $N_\uparrow$ and $N_\downarrow$.

\subsection{Hubbard-Stratonovich transformation} \label{HS_trans}
The Gibbs operator $e^{-\beta \hat{H}}$ at inverse temperature $\beta=1/k_BT$ can be thought of as the propagator in imaginary-time $\beta$. Decomposing the Hamiltonian into $\hat{H} = \hat{K} + \hat{V}$, where $\hat K$ is the kinetic energy operator and $\hat V$ is the two-body interaction, we use a symmetric Trotter decomposition
\begin{equation}
e^{-\beta \hat{H}} = [e^{-\Delta \beta \hat{K}/2}e^{-\Delta \beta \hat{V}}e^{-\Delta \beta \hat{K}/2}]^{N_{\tau}} + O(\Delta \beta^{2}) \;,
\end{equation}
where we have divided the imaginary time into $N_{\tau}=\beta/{\Delta \beta}$ discrete time slices of length $\Delta \beta$.  The number operator, $\hat{n}_{\bold{x}}$, at site $\bold x$ is given by $\hat{n}_{\bold{x}}=\hat{n}_{\bold{x},\uparrow} + \hat{n}_{\bold{x},\downarrow}$. For fermions, each interaction term in Eq.~(\ref{Hamiltonian}) can be written as  $g\hat{n}_{\bold{x},\downarrow}\hat{n}_{\bold{x},\uparrow}=\frac{1}{2}g(\hat{n}_{\bold{x}}^{2}-\hat{n}_{\bold{x}})$.  Completing the squares, we obtain the Hubbard-Stratonovich transformation~\cite{Hubbard1959,Stratonovich1957},
\begin{equation}
e^{-\Delta\beta g\hat{n}_{\bold{x}}^{2}/2}=\sqrt{\frac{\Delta\beta |g|}{2\pi }}\int_{-\infty }^{\infty }d\sigma_{\bold x} e^{-\Delta\beta |g| \sigma_{\bold x} ^{2}/2} e^{-\Delta\beta s g \sigma_{\bold x} \hat{n}_{\bold{x}}} \;,
\end{equation}
where $s=1$ for $g<0$ and $s=i$ for $g>0$. This is repeated for each of the $N_{\tau}$ time slices and $N_{L}^{3}$ lattice sites, and leads, in the limit $\Delta\beta\to 0$, to the functional integral form
\begin{equation}\label{HS} 
e^{-\beta \hat{H}} = \int D[\sigma ]G_{\sigma }\hat{U}_{\sigma } \;.
\end{equation}
The integration in Eq.~(\ref{HS}) is over auxiliary fields $\sigma_{\bold{x}}(\tau_n)$ defined for each lattice site $\bold{x}$ and imaginary time slice $\tau_n=n\Delta\beta$ ($n=0, 1,\ldots, N_\tau$), and the integration measure $D[\sigma]=  \prod_{{\bf x}, n}\left[d\sigma_{\bf x} (\tau_n)  \sqrt{{{\Delta}\beta |g| /{2\pi}}}\right]$.  $\hat{U}_{\sigma}$ is a time-ordered product for the propagator of non-interacting fermions in the imaginary-time dependent fields $\sigma_{\bf x}(\tau)$, given explicitly by 
\begin{equation}
\hat{U}_{\sigma}=\prod_{n}e^{-\Delta \beta \hat K/2} e^{-\Delta \beta \hat{h}_{\sigma}(\tau_n)} e^{-\Delta \beta \hat K/2}\;,
\end{equation}
where $\hat{h}_{\sigma}(\tau_n)=g\sum_{\bf x} \sigma_{\bf x}(\tau_n)\hat{n}_{\bf x} - g\hat{N}/2$ and $G_\sigma=e^{-\sum_n \Delta\beta |g|\sigma_{\bf x}(\tau_n)/2}$ is a Gaussian weight. The operator $\hat{U}_{\sigma}$ describes the many-particle propagator of the non-interacting spin-balanced system of fermions in external time-dependent one-body auxiliary fields $\sigma_{\bold{x}}(\tau)$.

\subsection{Thermal expectation values of observables and Gaussian quadratures} 
 The thermal expectation value $\langle \hat{O}\rangle$ of an observable is given by
\begin{equation} 
\langle \hat{O} \rangle=\frac{\textrm{Tr}(\hat{O}e^{-\beta \hat{H}})}{\textrm{Tr}(e^{-\beta \hat{H}})}=\frac{\int D[\sigma] \langle \hat{O} \rangle _{\sigma}W_{\sigma}\Phi_{\sigma} }{\int D[\sigma]W_{\sigma}\Phi_{\sigma}} \;,
\end{equation}
where $W_{\sigma}=G_{\sigma}|\textrm{Tr}\hat{U}_{\sigma}|$ is a positive-definite weight function, $\Phi_{\sigma}=\textrm{Tr}\hat{U}_{\sigma}/|\textrm{Tr}\hat{U}_{\sigma}|$ is the sign function, and $\langle \hat{O} \rangle_\sigma=\textrm{Tr}(\hat{O}\hat{U}_\sigma)/\textrm{Tr}\hat{U}_{\sigma}$ is the expectation value for a single configuration $\sigma$ of the auxiliary fields.  At each site $\bold x$ and time slice, we approximate the integral over $\sigma_{\bold x}$ by Gaussian quadrature
\begin{multline}
\sqrt{\frac{\Delta\beta |g| }{2\pi}}\int_{-\infty}^{\infty} d\sigma_{\bold x}  e^{-\Delta \beta |g| \sigma_{\bold x}^{2}/2}e^{-\Delta\beta s g \sigma_{\bold x}  \hat{n}_{\bold x}}\\
=\sum_{j=-1,0,1} w_{j} e^{-\Delta\beta s g \sigma_{\bold{x}, j} \hat{n}_{\bold x}}+O(\Delta\beta^{3})\;,
\end{multline}
where $\sigma_{\bold x, \pm 1}=\pm \sqrt{3/(\Delta\beta |g|)}$ and $\sigma_{\bold x,0}=0$.  The weights $w_{j}$ are given by $w_{\pm 1}=1/6$ and $w_{0}=2/3$. We sample the discrete auxiliary fields $\sigma_{\bold x, j}(\tau)$ according to the distribution $W_{\sigma}$ using the Metropolis algorithm~\cite{Metropolis1953}.\\

\subsection{Particle-number projection}  
The particle-number projection operator for $N$ particles and $N_{s}$ single-particle states can be written as a discrete Fourier transform~\cite{Koonin1994}
\begin{equation}\label{particle-projection}
\hat{P}_{N}=\frac{e^{-\beta \mu N}}{N_{s}}\sum_{m=1}^{N_{s}}e^{-i\varphi _{m}N}e^{(\beta \mu +i\varphi _{m})\hat{N}} \;,
\end{equation}
where $\varphi _{m}=\frac{2\pi m}{N_{s}}$ ($m=1,\ldots,N_s$)  are quadrature points, and $\mu$ is a chemical potential introduced to ensure the numerical stability of the Fourier sum.  $\mu$ is chosen so as to reproduce the correct number of particles $N$ in a grand-canonical average in the ensemble $U_\sigma$, i.e., ${\rm Tr}(\hat{N}\hat U_\sigma e^{\beta \mu \hat N})/{\rm Tr}(\hat U_\sigma e^{\beta\mu \hat N})=N$. 
We can calculate grand-canonical traces over the many-particle Fock space by using the matrix representation $\bold U_\sigma$ of the many-particle propagator $\hat U_\sigma$ in the single-particle space. For example 
\begin{equation}\label{partition-function}
\textrm{Tr}[e^{(\beta \mu +i\varphi_{m})\hat{N}}\hat{U}_\sigma]=\textrm{det}[\unit +e^{(\beta \mu +i\varphi _{m})}\bold{U}_\sigma] \;,
\end{equation}
where $\unit$ is the identity matrix in the single-particle space. 
Thus the canonical trace of $\hat U_\sigma$ at fixed particle number $N$ is given by
\begin{equation}
\textrm{Tr}_N \hat{U}_\sigma=\frac{e^{-\beta \mu N}}{N_s}\sum_{m=1}^{N_s}e^{-i \varphi _m N} \det [\unit + e^{(i \varphi _{m} +\beta \mu)}\bold{U}_{\sigma}] \;.
\end{equation}

In our AFMC calculations, we choose the single-particle basis to be the single-particle states with good quasi-momentum  $\hbar \bold{k}$ and spin projection $\sigma$. The single-particle propagator matrices $\bold{U}_{\sigma}$ are then evaluated in momentum space. When calculating the short-time propagator with the linearized interaction term, we obtain a speedup for the local interaction in coordinate space using a fast Fourier transform approach~\cite{Batrouni1986, Davies1988, Bulgac2006, Drut2013}. All observables are calculated with two number projections for fixed $N_{\uparrow}$ and fixed $N_{\downarrow}$ with the exception of the spin susceptibility. The spin susceptibility uses only one number projection for total particle number $N=N_{\uparrow}+N_{\downarrow}$ as discussed in Section~\ref{sec:spin_susceptibility}.

\subsection{Monte Carlo sign}

The spin-balanced Fermi gas with an attractive contact interaction has a good Monte Carlo sign, i.e., the canonical trace of $\hat U_\sigma$ is positive for each sample $\sigma$ of the auxiliary fields.  To show this, we note that in general
\begin{equation}\label{eq:factorization}
{\rm Tr}_{N_\uparrow, N_\downarrow}  \hat U_\sigma = {\rm Tr}_{N_\uparrow} \hat U_\sigma^\uparrow\; {\rm Tr}_{N_\downarrow}  \hat U_\sigma^\downarrow \;.
\end{equation}  
For an attractive interaction, the propagators $\hat U_\sigma^\uparrow$ and $\hat U_\sigma^\downarrow$ are invariant under time reversal and both particle-number projected traces on the right-hand-side of Eq.~(\ref{eq:factorization}) are real.  For the spin-balanced system $N_\uparrow=N_\downarrow$,  we have ${\rm Tr}_{N_\uparrow, N_\downarrow}  \hat U_\sigma =\left( {\rm Tr}_{N_\uparrow} \hat U_\sigma^\uparrow\right)^2$ and thus the canonical trace is positive for any sample $\sigma$. 

\subsection{Numerical stabilization}\label{sec:stabilization}
The calculation of the propagator matrix $\bold U_\sigma$ requires the multiplication of $N_{\tau}=\beta/\Delta \beta$ matrices. In this section we use the simplified  notation $U\equiv \bold{U}_{\sigma} =\prod_{i=1}^{N_{\tau}}U_{i}$ where $U_{i}$ denotes the $i^{th}$ imaginary-time step propagator. 
The condition number of $U$ is defined by $\kappa = ||U||||U^{-1}||$, where $||U||$ is a matrix norm~\cite{Meyer2000, Golub1996} such as the largest eigenvalue of $U$ or the Frobenius norm,
\begin{equation}
||U||=\sqrt{\sum_{i=1}^{N_s}\sum_{j=1}^{N_s}|U_{ij}|^2}\;.
\end{equation}
As $N_{\tau}$ becomes large and we approach low temperatures (large $\beta$), the condition number $\kappa$ can become very large and the product $U$ is ill-conditioned, i.e., we lose relevant information regarding the separation of large and small scales.

This problem is resolved in AFMC calculations by decomposing the propagator into a singular-value decomposition (SVD)~\cite{Loh1989, Loh1992, Loh2005, Gubernatis2016}, or a QDR decomposition, in order to keep the separation of scales in a diagonal matrix and avoid the loss of information from mixing scales.  In both cases, the matrix $U$ is decomposed into a product $U=ADB$, where $D$ is a diagonal matrix with the different scales and $A$ and $B$ are well-conditioned matrices~\cite{Loh1992, Koonin1997, Gilbreth2015, Gubernatis2016}. This can be described schematically by
\begin{equation} \label{eq:SVD}
  U = \left(\begin{array}{cccc}
     \scriptscriptstyle{X} & \scriptscriptstyle{X} & \scriptscriptstyle{X} & \scriptscriptstyle{X}\\
     \scriptscriptstyle{X} & \scriptscriptstyle{X} & \scriptscriptstyle{X} & \scriptscriptstyle{X}\\
     \scriptscriptstyle{X} & \scriptscriptstyle{X} & \scriptscriptstyle{X} & \scriptscriptstyle{X}\\
     \scriptscriptstyle{X} & \scriptscriptstyle{X} & \scriptscriptstyle{X} & \scriptscriptstyle{X}
\end{array}\right)  \left(\begin{array}{cccc}
     X & & & \\
     & \scriptstyle{X} & & \\
     & & \scriptscriptstyle{X} & \\
     & & & \scriptscriptstyle{X}
\end{array}\right) \left(\begin{array}{cccc}
     \scriptscriptstyle{X} & \scriptscriptstyle{X} & \scriptscriptstyle{X} & \scriptscriptstyle{X}\\
     \scriptscriptstyle{X} & \scriptscriptstyle{X} & \scriptscriptstyle{X} & \scriptscriptstyle{X}\\
     \scriptscriptstyle{X} & \scriptscriptstyle{X} & \scriptscriptstyle{X} & \scriptscriptstyle{X}\\
     \scriptscriptstyle{X} & \scriptscriptstyle{X} & \scriptscriptstyle{X} & \scriptscriptstyle{X}
\end{array}\right) \;,
\end{equation}
where the size of the characters represents the numerical scales. 

In our implementation, we use a QDR decomposition, i.e., $U= QDR$, where $Q$ is a unitary matrix, $R$ is upper right triangular, and $D$ is diagonal. To diagonalize $U$, one cannot multiply out the factors $QDR$ in this order without losing information contained in the smaller scales of $D$. Using the standard method to calculate $\textrm{Tr}_{N}(\hat{U}_{\sigma})$ stabily, which involves a QDR decomposition for each Fourier sum component in the canonical trace, would lead to scaling with system size of $O({N_s}^4)$ (as opposed to $O({N_s}^3)$ for grand canonical calculations). This is costly as the number of single-particle states $N_s$ grows larger. Instead, one can diagonalize the matrix $S = DRQ$ obtained from $U$ by a similarity transformation, and from this calculate number-projected observables. This procedure reduces the number of operations required for numerically stabilized canonical-ensemble calculations from $O(N_s^4)$ (without diagonalization) to $O (N_s^3)$~\cite{Gilbreth2015}. This method is used in our code. Recently, a new algorithm in the canonical ensemble was introduced in Ref.~\cite{Shen2023} using a recursive method which was shown to be numerically stable. It will be interesting to compare both methods in future work.

\subsection{Model space truncation}\label{sec:trunc}

In our simulations we use the model space truncation method first introduced in Ref.~\cite{Jensen2019} and presented in detail in Ref.~\cite{Gilbreth2021}. This allows for substantial savings in computational time. 

The main idea of the method is to reduce the dimension of the one-body propagator in the decomposition $U=QDR$ from $N_s$ to $N_{\rm occ}$, where $N_{\rm occ}$, the number of significantly occupied single-particle states,  is of the order of particle number $N$.  The entries of the diagonal matrix $D$ can be sorted in an appropriate descending order and decompose  $D = D_{\rm occ} + D_{\varepsilon}$, where $D_{\rm occ}$ is obtained by zeroing out the diagonal elements for row $i>N_{\rm occ}$ (and the discarded terms are in $D_{\varepsilon}$).  This leads to 
\begin{equation} \label{truncation}
  U = Q D_{\rm occ} R + Q D_{\varepsilon} R = U_{\rm occ} + E \;,
\end{equation}
where $U_{\rm occ} = Q D_{\rm occ} R$ represents the occupied part of $U$ and $E = Q D_{\varepsilon} R$ is a small perturbation.  All  observables are then calculated using the reduced-rank matrix $U_{\rm occ}$.  The error made is represented by the matrix $E$ with norm $||E||$. Our algorithm implementation uses the Frobenius norm and we choose $N_{\rm occ}$ such that the error matrix $E$ satisfies $||E||<\varepsilon$, where $\varepsilon$ is a given small parameter. Ref.~\cite{Gilbreth2021} provides an error analysis and further details of the truncation method. 

The benefit of the truncation is to replace the original $O(N_{\tau} N_s^3)$ time complexity of computing the partition function for a given field configuration $\sigma$ of Eq.~\ref{partition-function} in the complete single-particle model space with $O(N_{\tau} N_s N_{\rm occ}^2)$ complexity. This reduction in complexity becomes particularly large at low temperatures where $N_{\rm occ}$ is comparable to particle number $N$.  Further optimizations to reduce the $O(N_{s}^3)$ complexity of the $QR$ decompositions were introduced in the work of Ref.~\cite{He2019_RC} and applied in grand-canonical simulations of the two-dimensional BCS-BEC crossover~\cite{He2022}. These additional optimizations will be useful for reaching even larger lattice sizes and, in particular, lower temperatures in future studies of the UFG.

\section{Results and analysis}\label{sec:Results}

In Ref.~\cite{Jensen2020a} we performed simulations for $N=20, 40, 80$ and $130$ particles for a filling factor of $\nu\approx 0.06$. The finite filling factor introduces a systematic error as discussed in Sec.~\ref{sec:AFMC_pseudogap}. In Ref.~\cite{Jensen2020b}, we eliminated this systematic error when calculating the temperature dependence of Tan's contact across the superfluid phase transition for $N=40, 66$ and $114$ particles. This was accomplished  by extrapolating to the continuum limit using results obtained for lattice sizes of $N_{L}^3=5^3,7^3,9^3,11^3,13^3$ and $15^3$. Here we extend the simulations of Ref.~\cite{Jensen2020b} to extrapolate to the continuum limit for several other thermodynamic observables including the condensate fraction, the energy-staggering pairing gap, the spin susceptibility, and the free energy-staggering pairing gap. We also performed simulations for $N=54$ particles for temperatures within the critical region to reduce the statistical error of our estimate for the critical temperature calculated by finite-size scaling. 

In our calculations, for each particle number $N$, temperature $T/T_{F}$, and lattice size $N_{L}^3$, we also carry out simulations with several (typically four) values of the time slice $\Delta \beta$ and extrapolate to the continuous imaginary-time limit $\Delta\beta \to 0$. 

In Appendix~\ref{app:extrapolations}, we discuss further details of the extrapolations to continuous time $\Delta \beta \to 0$ and to the continuum limit  $\nu \to 0$ and present typical extrapolation plots for the  condensate fraction and free energy staggering pairing gap. 

\subsection{Condensate Fraction}

We determine the superfluid critical temperature $T_c$ by analyzing the condensate fraction $n$ as a function of temperature $T/T_{F}$ and particle number $N$ in the continuum limit of the lattice model.  We define the condensate fraction $n$ by
\begin{equation}
n=\frac{\lambda}{(N/2)} \;,
\end{equation}
where $\lambda$ is the largest eigenvalue of the two-body density matrix 
$\langle\hat{\psi}^{\dagger}_{\mathbf{k}_1,\uparrow}\hat{\psi}^{\dagger}_{\mathbf{k}_2,\downarrow}\hat{\psi}_{\mathbf{k}_3,\downarrow}\hat{\psi}_{\mathbf{k}_4,\uparrow}\rangle$, and $N/2$ is the maximal number of pairs.  For fermions, $\lambda$ can be shown to satisfy 
\begin{equation}
\lambda\leq {N(M-N/2+1)}/(2M) \leq N/2 \;,
\end{equation}
where $M=N_L^3$ is the number of lattice points. Having a large eigenvalue $\lambda$ that scales with the system size is referred to as off-diagonal long-range order (ODLRO)~\cite{Yang1962}.


\begin{figure*}[tbh]
\includegraphics[scale=1.4]{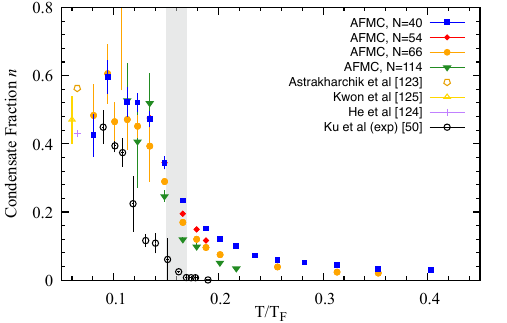}
\caption{AFMC condensate fraction $n$ in the continuum limit for $N=40$ (solid blue squares), $N=66$ (solid orange circles), and $N=114$ (solid green down triangles) particles.  We also include results for $N=54$ particles (solid red diamonds)  in the critical regime.  For reference we  show the $T=0$ quantum Monte Carlo result of Ref.~\cite{Astrakharchik2005} (open orange pentagon), the $T=0$ QMC result of Ref.~\cite{He2019} (purple cross), the recent low-T experimental result of Ref.~\cite{Kwon2020} (open up yellow triangle), and the finite temperature rapid ramp experimental result of Ref.~\cite{Ku2012} (black open circles). The grey band represents the critical temperature $T_c=0.16(1)\,T_{F}$ determined from finite-size scaling.}\label{ODLRO-continuum}
\end{figure*}

In Fig.~\ref{ODLRO-continuum} we show the condensate fraction $n$ (after the continuous time and continuum extrapolations discussed in Appendix \ref{ext-n}) as a function of temperature $T/T_F$ for $N=40$ (solid blue squares), $N=66$ (solid orange circles), and $N=114$ (solid green down triangles) particles. We also show results for $N=54$ particles (solid red diamonds) in the vicinity of the critical temperature.  There is a noticeable difference between the results of Ref.~\cite{Jensen2020a} and our current continuum results. This is expected and demonstrates the importance of the continuum limit extrapolations in obtaining reliable results from lattice simulations.  We also show the $T=0$ quantum Monte Carlo result of Ref.~\cite{Astrakharchik2005} (open orange pentagon) where a diffusion Monte Carlo calculation was carried out for the full BCS to BEC crossover, and the $T=0$ quantum Monte Carlo (QMC) result of Ref.~\cite{He2019} (purple cross) where a more complex lattice interaction is introduced to accelerate the convergence to the continuum limit. We compare with the recent low-temperature Josephson current experimental results of Ref.~\cite{Kwon2020} (open up yellow triangle) and the finite temperature rapid ramp experimental result of Ref.~\cite{Ku2012} (open black circles). 

The grey band represents our estimate for the critical temperature $T_c=0.16(1)T_{F}$ from a finite-size scaling analysis (see Sec.~\ref{FSS}). We observe that the condensate fraction decreases with particle number $N$ in the normal phase as expected when the thermodynamic limit is approached in a finite-size system, but it does not show a noticeable trend with $N$ below the critical temperature. Our results for the condensate fraction are larger than the rapid ramp experimental results of Ref.~\cite{Ku2012} for all temperatures.  Comparing our $T<<T_c$ results with the $T=0$ results of other works,  we find good agreement with the diffusion Monte Carlo result of Ref.~\cite{Astrakharchik2005}  (open orange pentagon) and with the experimental result of Ref.~\cite{Kwon2020} (open up yellow triangle). Our low-temperature results are slightly higher than the improved lattice interaction result of Ref.~\cite{He2019} (purple cross).

\subsection{Finite-size scaling and critical temperature}\label{FSS}

To determine the superfluid critical temperature $T_c$ we apply a finite-size scaling analysis to the condensate fraction $n$ for $N=40, 54, 66$ and $N=114$ (the $N=54$ simulations were performed only in the critical regime to improve our estimate of $T_c$).  In Refs.~\cite{Burovski2006, Burovski2006-2, Akkineni2007, Goulko2010} a finite-size scaling analysis, discussed in Ref.~\cite{Binder1981},  was applied to the 3D XY phase transition of the UFG to determine the critical temperature. The critical behavior of the continuum limit condensate fraction $n(L,T)$ is captured by the scaling ansatz 
\begin{equation}\label{Eq.scaling_relation}
L^{1+\eta}n(L,T)=f(x)\ [ 1 + C L^{-\omega} + ... \ ] \;,
\end{equation}
where $x=(L/\xi)^{1/\nu_c}$, $\xi$ the correlation length, $f(x)$ is a universal scaling function, and $C$ is a non-universal coefficient of the term for the leading correction to scaling with an exponent $\omega$. The exponents of the $U(1)$ universality class have previously been calculated with $\nu_c\simeq 0.67$, $\eta\simeq 0.038$ with $\omega \approx 0.8$ for the exponent of the leading irrelevant field~\cite{Guida1998, Campostrini2006, Pelissetto2002}. 

\begin{figure}[b]
\includegraphics[scale=1.05]{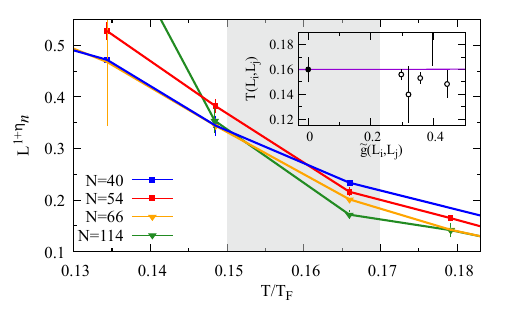}
\caption{Continuum limit condensate fraction $n(L,T)$ scaled by $L^{1+\eta}$ as a function of temperature $T/T_F$ for $N=40$ (solid blue squares), $N=54$ (solid red diamonds), $N=66$ (solid orange down triangles), and $N=114$ (solid green down triangles).  Inset: the crossing temperatures $T(L_{i},L_{j})$ as a function of $\tilde g (L_{i},L_{j})$ (discussed in the main text) and the fit (straight purple line) to extract the  critical temperature $T_c$. We find $T_c=0.16(1)T_{F}$ (solid black circle) indicated by the vertical grey band in the main figure.}\label{ODLRO-FSS}
\end{figure}

In Fig.~\ref{ODLRO-FSS} we plot the condensate fraction in the continuum limit $n(L,T)$ scaled by $L^{1+\eta}$ as a function of $T/T_F$ for the different particle numbers. $L$ does not represent a simulated lattice size but the size of the physical system with $N$ particles. We arbitrarily choose $L'=1$ for $N'=40$ particles in Fig.~\ref{ODLRO-continuum}(b) and the $L$ values for the remaining particle numbers are determined through the relation $N'/L'^3=N/L^3$. If the corrections to the scaling ansatz are vanishing the crossings of the curves for different particle number $N$ scaled by $L^{1+\eta}$ will occur at the superfluid critical temperature $T_c$. However, the corrections to the scaling relation can be significant for the finite particle numbers simulated in the current work. We follow Ref.~\cite{Burovski2006, Burovski2006-2, Goulko2010} and expand $f(x)$ to leading order ($f(x)$ is analytic at $x=0$), where the correlation length is $\xi \propto |t|^{-\nu_{c}}$ in the critical regime where the reduced temperature is $t=(T-T_c)/T_c$. We then obtain the relation between the crossing temperature for system sizes $L_{i}$ and $L_{j}$, denoted by $T(L_{i},L_{j})$, and the thermodynamic critical temperature $T_c$ 
\begin{equation}
|T(L_{i},L_{j})-T_{c}| = \kappa g(L_{i},L_{j})
\end{equation}
with
\begin{equation}\label{Eq.g_full}
g(L_{i},L_{j})=\frac{\ [ L_{i}^{-\omega}-L_{j}^{-\omega}\ ] }{\ [ (L_{j}^{1/\nu_{c}}-L_{i}^{1/\nu_{c}})+C(L_{j}^{1/\nu_{c} - \omega}-L_{i}^{1/\nu_{c}-\omega})\ ]}\;.
\end{equation}
Here $\kappa$ absorbs constants in the expansion of the universal function $f(x)$.  Refs.~\cite{Burovski2006,Burovski2006-2} dropped the second term in the denominator of Eq.~(\ref{Eq.g_full}) to obtain a simplified relation 
\begin{equation}
\widetilde{g}(L_{i},L_{j})=\frac{\ [ L_{i}^{-\omega}-L_{j}^{-\omega}\ ] }{\ [ (L_{j}^{1/\nu_{c}}-L_{i}^{1/\nu_{c}})\ ]}.
\end{equation}

In this work the data is not precise enough to perform the analysis of  Ref~\cite{Goulko2010} and instead we extract $T_c$ with the simplified crossings function $\widetilde{g}(L_{i},L_{j})$ . In the inset of Fig.~\ref{ODLRO-FSS}  we show the crossing temperatures $T(L_{i},L_{j})$ (open black circles) as a function of $\widetilde{g}(L_{i},L_{j})$ for $N=40,54,66$ and $114$ particles where the crossings are determined from straight line fits to the Monte Carlo data for each particle number $N$. We have performed a generalized least square analysis to determine the critical temperature $T_c=0.16(1)\, T_{F}$ (solid black circle). This is larger  than the estimate $T_c = 0.130(15) \,T_F$ of Ref.~\cite{Jensen2020a}, where simulations were performed at a finite filling factor of $\nu\approx 0.06$. We observe that the critical temperature extracted from the AFMC data in the continuum limit is closer to the experimental value $T_c=0.167(13) T_F$ of Ref.~\cite{Ku2012} and to the continuum limit diagrammatic Monte Carlo estimates $T_{c}=0.152(7) \, T_{F}$ of Ref.~\cite{Burovski2006} and $T_{c}=0.173(6)\, T_{F}$ of Ref.~\cite{Goulko2010}. The path integral Monte Carlo work of Ref.~\cite{Akkineni2007} found a higher critical temperature $T_c = 0.245 \,T_F$. There are several other theoretical estimates for the critical temperature in Refs.~\cite{Bulgac2006, Bulgac2008, Lee2006, Nishida2007, Haussmann2007, Abe2009, Akkineni2007, Floerchinger2010, Boettcher2014}. Experimental results for $T_c$ were obtained in the works of Refs.~\cite{Ku2012, Navon2009, Horikoshi2010}. For reference we list the critical temperature estimates of various experiments and theoretical calculations in Table ~\ref{Table:FSS}.
\begin{table}
\caption{Estimates of the critical temperature $T_c$ (in units of the Fermi temperature $T_F$)}\label{Table:FSS}
\centering
\begin{tabular}{c c c c}
\hline\hline
Method  & $T_c (T_F)$ & error \\ [0.5ex] 
\hline
MIT experiment~\cite{Ku2012}&0.167&0.013\\
experiment~\cite{Navon2009}&0.157&0.015\\
experiment~\cite{Horikoshi2010}& 0.17& 0.01\\
lattice DiagMC~\cite{Burovski2006}&0.152&0.007\\
lattice DiagMC~\cite{Goulko2010}&0.173&0.006\\
lattice AFMC~\cite{Richie2020} &0.16&0.02\\
lattice AFMC~\cite{Bulgac2008} &$\lesssim$~0.15&0.01\\
lattice AFMC~\cite{Bulgac2006} & 0.23 & 0.02\\
low-density neutron~\cite{Abe2009} & 0.189 & 0.012\\
RPIMC~\cite{Akkineni2007} & $\approx0.245$ & \\
lattice Monte Carlo~\cite{Lee2006} & $<$~0.14& \\
FRG~\cite{Floerchinger2010}& 0.248 & \\
$\epsilon$-expansion matching from d=2,4~\cite{Nishida2007} & 0.183 & 0.014\\
Luttinger-Ward~\cite{Haussmann2007} & 0.183 & 0.014\\
lattice AFMC (this work) & 0.16 & 0.01\\[1ex]
\hline
\end{tabular}
\label{table:gap}
\end{table}

\subsection{Energy-staggering pairing gap}
 
 
  \begin{figure}[bth]
\centering
\includegraphics[scale=1.02]{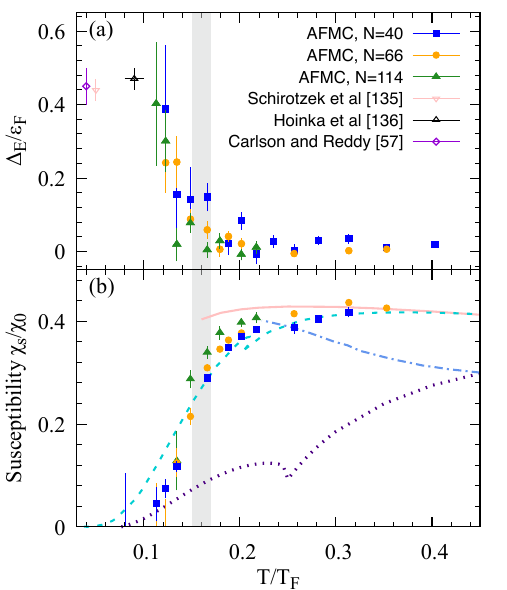}
\caption{(a) AFMC energy-staggering pairing gap $\Delta_E$ (in units of the Fermi energy $\varepsilon_F$) for $N=40$ (solid blue squares), $N=66$ (solid orange circles), and $N=114$ (solid green down triangles) particles as a function of $T/T_F$. For reference, we show the low-T experimental result of Ref.~\cite{Schirotzek2008} (open down pink triangle), the low-T experimental result of Ref.~\cite{Hoinka2017} (open up black triangle), and the low-T hybrid QMC experimental result of Refs.~\cite{Carlson2008} (open purple diamond). The AFMC energy-staggering pairing gap vanishes above a temperature of $T^*\approx 0.2 \,T_F$ for the largest particle number simulated.  The gray band is our estimate of the critical temperature $T_c$ (in units of $T_F$).  (b) The AFMC spin susceptibility $\chi_s$ (in units of $\chi_0$) for different particle numbers as a function of $T/T_F$. The convention of the symbols for different particle numbers is as in panel (a). For comparison we also show the spin susceptibility calculated in the Luttinger-Ward theory~\cite{Enss2012} (solid orange line), the $t$-matrix results of Ref.~\cite{Palestini2012} (dotted line), the extended $T$-matrix result of Refs.~\cite{Tajima2014,Tajima2016} (dashed line), and the self-consistent Nozieres and Schmitt-Rink results of Ref.~\cite{Pantel2014} (dashed-dotted line).}\label{fig:pairing}
\end{figure}

\begin{figure*}[bth]
\begin{center}
\includegraphics[width=2\columnwidth]{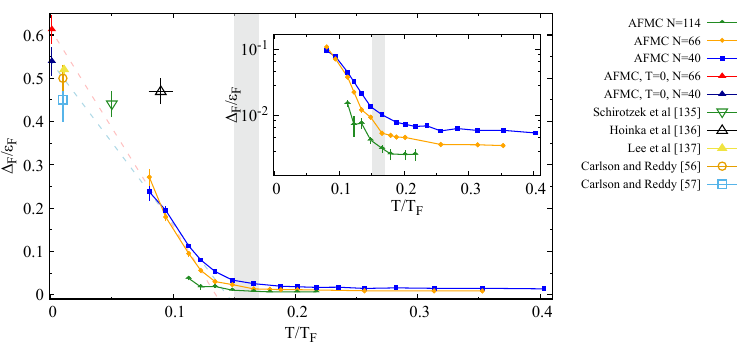}
\end{center}
\caption{AFMC free energy gap $\Delta_F$ (in units of the Fermi energy $\varepsilon_{F}$) for $N=40$ (solid blue squares), $N=66$ (solid orange diamonds), and $N=114$ particles (solid green up triangles). We also show the linear extrapolations in $T/T_{F}$ for $N=40$ (blue dashed line) and $N=66$ (pink dashed line), and the $T=0$ energy-staggering pairing gap $\Delta_{E}/\varepsilon_{F}$ from the $N=40$  (solid blue up triangle) and $N=66$ (solid red up triangle) extrapolations. The gray band shows the critical temperature $T_{c}=0.16(1)T_{F}$ determined from a finite-size scaling analysis.  For reference, we show the low-temperature gap from the QMC results of Refs.~\cite{Carlson2005} (open orange circle) and \cite{Carlson2008} (open blue square), the lattice estimator QMC result of Ref.~\cite{Lee2010} (solid yellow up triangle), and the experimental values of Ref.~\cite{Schirotzek2008} (open down green triangle) and Ref.~\cite{Hoinka2017} (open up black triangle). Inset: The free energy gap $\Delta_F$ on a log-linear scale, showing more clearly the increase in $\Delta_F$ as the temperature decreases below a certain temperature which we associate with a pairing temperature scale $T^*$.  For the largest particle number $N=114$ we estimate an upper bound of $T^{*} \le 0.2 \, T_F$.}\label{fig:FGAP}
\end{figure*}

 A signature of a pseudogap regime is the non-vanishing of a pairing gap above the critical temperature. We define a model-independent energy-staggering pairing gap $\Delta_E$ at temperature $T$  for an even number of particles $N$ by 
\begin{multline}\label{pairing-gap}
\Delta_{E}= [2E(N/2+1,N/2)- E(N/2+1,N/2+1) \\-E(N/2,N/2)]/2 \;,
\end{multline}
where $E(N_{\uparrow},N_{\downarrow})$ is the thermal energy for $N_{\uparrow}$ spin-up particles and $N_{\downarrow}$ spin-down particles.  This observable was first calculated for the UFG at $T=0$ in Ref.~\cite{Carlson2003} and at finite temperature in the works of Refs.~\cite{Jensen2019, Jensen2020a}. 

The calculation of $\Delta_E$ at finite temperature requires the use of the canonical ensemble of fixed particle numbers. To this end, we use the particle-number reprojection method of Ref.~\cite{Alhassid1999}. The expectation value of the Hamiltonian $\hat{H}$ for fixed particle numbers $N_{\uparrow}$ and $N_{\downarrow}$ is given by
\begin{equation} 
\langle \hat{H} \rangle_{N_{\uparrow}, N_{\downarrow}}=\frac{\textrm{Tr}_{N_{\uparrow},N_{\downarrow}}(\hat{H}e^{-\beta \hat{H}})}{\textrm{Tr}_{N_{\uparrow},N_{\downarrow}}(e^{-\beta \hat{H}})}=\frac{\int D[\sigma] \langle \hat{H} \rangle _{N_{\uparrow}, N_{\downarrow},\sigma}W_{\sigma}\Phi_{\sigma} }{\int D[\sigma]W_{\sigma}\Phi_{\sigma}}
\end{equation}
where 
\begin{equation}\label{weight}
W_{\sigma}=G_{\sigma}|\textrm{Tr}_{N_{\uparrow},N_{\downarrow}} \hat{U}_{\sigma} | 
\end{equation}
 is a positive definite weight function used for the Monte Carlo sampling, $\Phi_{\sigma}=\textrm{Tr}_{N_{\uparrow},N_{\downarrow}}(\hat{U}_{\sigma})/|\textrm{Tr}_{N_{\uparrow},N_{\downarrow}}(\hat{U}_{\sigma})|$ is the Monte Carlo sign, and 
\begin{equation}
\langle \hat{H} \rangle_{N_{\uparrow}, N_{\downarrow},\sigma} = {\textrm{Tr}_{N_{\uparrow},N_{\downarrow}}(\hat{H}\hat{U}_\sigma) \over \textrm{Tr}_{N_{\uparrow},N_{\downarrow}}(\hat{U}_{\sigma}) }\;,
\end{equation} 
is the expectation value of $H$ at fixed spin-up ad spin-down particle numbers for a given configuration $\sigma$ of the auxiliary fields.  We use exact particle-projection operators given by Eq.~(\ref{particle-projection}).  

To calculate  $\langle \hat{H} \rangle_{N'_{\uparrow}, N'_{\downarrow}}$ for different particle numbers $N'_{\uparrow}, N'_{\downarrow}$, we use the original Monte Carlo sampling for $N_{\uparrow},N_{\downarrow}$ particles  but reproject on the new particle numbers~\cite{Alhassid1999}.
\begin{equation} 
\langle \hat{H} \rangle_{N'_{\uparrow}, N'_{\downarrow}}= { \left\langle \langle \hat H \rangle_{N'_\uparrow, N'_\downarrow,\sigma} \frac{\textrm{Tr}_{N'_\uparrow, N'_\downarrow} \hat U_\sigma} {{\textrm{Tr}_{N_\uparrow,N_\downarrow}\hat U_\sigma}} \Phi_\sigma\right\rangle_W \over  \left\langle \frac{\textrm{Tr}_{N'_\uparrow, N'_\downarrow}\hat U_\sigma} {{\textrm{Tr}_{N_\uparrow,N_\downarrow} \hat U_\sigma}} \Phi_\sigma\right\rangle_W}  \;,
\end{equation}
where we have introduced the notation
\begin{equation}
\langle X_\sigma \rangle_W \equiv {{\int D[\sigma] W_\sigma X_\sigma \over
\int  D[\sigma] W_\sigma}} \;.
\end{equation} 

In Fig.~\ref{fig:pairing}(a) we show the continuum limit energy-staggering pairing gap $\Delta_E$ (measured in units of the Fermi energy $\varepsilon_F$) as a function of temperature $T/T_F$  for $N=40, 66$ and $114$ particles. We observe that for the largest particle number ($N=114$), the pairing gap vanishes above a temperature of $T^* \approx 0.2 \,T_F$. This suggests that the pseudogap regime, in which the pairing gap does not vanish above $T_c$ is considerably narrower than previously believed. 

\subsection{Spin susceptibility} \label{sec:spin_susceptibility}

The spin susceptibility $\chi_s$ is suppressed by pairing correlations and is another signature of a pseudogap regime. It is given by
\begin{equation} \label{susceptibility}
\chi_{s}=\frac{\beta}{V}\langle(\hat{N}_{\uparrow}-\hat{N}_{\downarrow})^{2}\rangle \;,
\end{equation}
where the expectation value on the r.h.s.~is calculated for the spin-balanced system $\langle \hat N_\uparrow \rangle  = \langle \hat N_\downarrow \rangle$ using only one particle projection on total number of particles $N=N_\uparrow + N_\downarrow$ in Eq.~(\ref{susceptibility}).

We calculated the spin susceptibility using the same Monte Carlo sampling at fixed numbers of spin-up and spin-down particles, i.e., according to $W_\sigma$ defined in Eq.~(\ref{weight}). We then have
\begin{equation} 
\chi_s=\frac{\beta}{V} {\left\langle \langle(\hat{N}_{\uparrow}-\hat{N}_{\downarrow})^2 \rangle_\sigma  \frac{\textrm{Tr}_{N}\hat U_\sigma} {{\textrm{Tr}_{N_\uparrow,N_\downarrow} \hat U_\sigma}} \Phi_\sigma \right\rangle \over  \left\langle  \frac{\textrm{Tr}_{N}\hat U_\sigma} {{\textrm{Tr}_{N_\uparrow,N_\downarrow} \hat U_\sigma}} \Phi_\sigma \right\rangle} \;,
\end{equation}
where $ \langle(\hat{N}_{\uparrow}-\hat{N}_{\downarrow})^2 \rangle_\sigma $ is the expectation value of $(\hat{N}_{\uparrow}-\hat{N}_{\downarrow})^2$ at fixed total particle number $N$ for a given field configuration $\sigma$. 

 In  Fig.~\ref{fig:pairing}(b) we show the continuum limit spin susceptibility $\chi_s$ (in units of the $T=0$ free Fermi gas susceptibility $\chi_0=3\rho/2\varepsilon_F$) as a function of $T/T_F$ for particle numbers $N=44, 66$ and $114$. We observe the strong suppression of $\chi_s$ below the critical temperature due to pairing correlations.  However, we also observe moderate suppression above $T_c$. While there is a clear particle number dependence and we have not reached the thermodynamic limit, we can put an upper bound on a spin gap temperature $T_s \approx 0.2\,T_F$ below which $\chi_s$ starts to become suppressed.   We observe that the value of the spin gap temperature is similar to that of $T^*$.
 
 Based on our results in Fig.~\ref{fig:pairing} for the energy-staggering pairing gap and spin susceptibility, we estimate the pseudogap regime above $T_c \approx 0.16\, T_F$ to be below $T^* \approx 0.2\, T_F$.
 
\subsection{Free energy gap and extraction of the $T=0$ gap}

The energy-staggering pairing gap is noisy, which makes the determination of an appropriate cutoff in filling factor $\nu$ for the continuum extrapolation rather difficult.  A much less noisy quantity to investigate pairing correlations is the free energy staggering pairing gap, which is defined by  a formula analogous to Eq.~(\ref{pairing-gap}) but with the thermal energy replaced by the free energy 
\begin{multline}
\Delta_{F}=[2F(N/2+1,N/2)- F(N/2+1,N/2+1)\\-F(N/2,N/2)]/2 \;.
\end{multline}
Here $F(N_{\uparrow},N_{\downarrow})= - k_BT \ln Z_{N_{\uparrow},N_{\downarrow}}$ and $ Z_{N_{\uparrow},N_{\downarrow}}$ are, respectively,  the free energy and partition function for $N_{\uparrow}$ spin-up particles and $N_{\downarrow}$ spin-down particles. 

Within the AFMC framework, this observable is calculated using the particle-number reprojection method.  We rewrite the free energy gap as
\begin{equation}
\Delta_{F} = -k_B T \ln \left(\frac{Z_{N/2,N/2+1}}{Z_{N/2,N/2}} \frac{Z_{N/2,N/2+1}}{Z_{N/2+1,N/2+1}}\right)  \;.
\end{equation}
Each partition function ratio can then be calculated using particle-number reprojection, e.g.,
\begin{equation}
\frac{Z_{N/2,N/2+1}}{Z_{N/2,N/2}}= \left\langle \frac{ {\rm Tr}_{N/2,N/2+1} U_\sigma} {{\rm Tr}_{N/2,N/2}  U_\sigma} \right\rangle_W \;,
\end{equation}
where  $W_\sigma= G_{\sigma}  {\rm Tr}_{N/2,N/2}  U_\sigma$ is positive definite for $N_{\uparrow} =N_{\downarrow} = N/2$.

In Fig.~\ref{fig:FGAP} we show the AFMC free energy gap $\Delta_F$ (in units of the Femi gas energy $\varepsilon_F$) as a function of temperature $T/T_F$ for  $N=40, 66$ and $114$ particles. The statistical errors on the free energy gap are small and this quantity can thus be determined more accurately than the energy-staggering pairing gap.  We see that the free energy gap is suppressed above the critical temperature $T_c$.  To see more clearly the behavior of $\Delta_F$ above $T_c$, we plot in the inset the free energy gap on a log-linear scale. We observe that as we decrease the temperature from above $T_c$, the free energy gap starts to increase below the pairing temperature scale $T^*$ that we identified in the energy-staggering pairing gap and the spin susceptibility in Fig.~\ref{fig:pairing}.  Extrapolation to the thermodynamic limit will be important in future works to determine whether the pseudogap temperature scale $T^*\approx 0.2 \, T_F$ is robust or will go down, reducing further the extent of the UFG pseudogap regime.

The free energy gap $\Delta_{F}$ can be used to obtain a more precise value for the $T=0$ energy-staggering pairing gap as follows.  We write $\Delta_{F}$ as $\Delta_{F}= \Delta_E  - T \Delta_S$, where $\Delta_S$ is the entropy gap. Both $\Delta_E$ and $\Delta_S$ saturate at low temperatures leading to a linear dependence of $\Delta_F$ on $T$ at low temperatures. In Fig.~\ref{fig:FGAP} we show linear extrapolations of the free energy gap to $T=0$ for $N=40$ (blue dashed line) and $N=66$ (pink dashed line). We then estimate the $T=0$ energy-staggering pairing gap $\Delta_{E}/\varepsilon_{F}$ by taking the average of the $N=40$ (solid blue up triangle) and $N=66$ (solid red up triangle) extrapolations to find $\Delta_E = 0.576 (24) \, \varepsilon_F$. We compare with the $T=0$ QMC results of Ref.~\cite{Carlson2005} (open orange circle), Ref.~\cite{Carlson2008} (open blue square), Ref.~\cite{Lee2010} (solid yellow up triangle), and the experimental results of Ref.~\cite{Schirotzek2008} (open down green triangle) and Ref.~\cite{Hoinka2017} (open up black triangle). For reference, in Table~\ref{table:gap} we list theoretical and experimental results for the $T=0$ pairing gap of the UFG.

\begin{table}[t!]
\caption{Estimates of the $T=0$ pairing gap}
\centering
\begin{tabular}{c c c c}
\hline\hline
Method  & $\xi$ & error \\ [0.5ex] 
\hline
MIT low-T experiment~\cite{Schirotzek2008}&0.44&0.03\\
Swinburne low-T experiment~\cite{Hoinka2017}&0.47&0.03\\
Ground-state fixed-node Monte Carlo~\cite{Chang2004} &0.594&0.024\\
Ground-state AFMC~\cite{Carlson2003} &0.55&0.05\\
Ground-state AFMC~\cite{Carlson2005} &0.50&0.03\\
QMC + MIT experiment~\cite{Carlson2008}&0.45&0.05\\
Luttinger-Ward~\cite{Haussmann2007}&0.46& \\
Functional renormalization group~\cite{Bartosch2009}&0.61& \\
Functional renormalization group~\cite{Floerchinger2010}&0.46&\\
Lattice quantum Monte Carlo~\cite{Lee2010}&0.52&0.01 \\
AFMC (this work) &0.576&0.024\\[1ex]
\hline
\end{tabular}
\label{table:gap}
\end{table}

\section{Conclusion and outlook}\label{sec:conclusions}

There has been substantial progress in recent years towards understanding the BCS-BEC crossover and the UFG in the two-species Fermi gas with short-range interactions . Experimental advances have provided an understanding of several observables for this strongly correlated system~\cite{Jin1999, Modugno2002, Truscott2001, Schreck2001, Granade2002, Hadzibabic2002, Jochim2003, Zwierlein2005, Navon2010, Ku2012, Carcy2019, Mukherjee2019}. However, from a theoretical perspective, there is no controlled analytic approach to the UFG. The current theoretical approaches that are controllable (and scale to study large system sizes) are quantum Monte Carlo methods. 

 In this work, we calculated the first lattice continuum limit of the condensate fraction at finite temperature for the homogeneous UFG.  The continuum limit extrapolations require calculations for large lattices and they were made possible through the development of canonical-ensemble AFMC optimizations in Refs.~\cite{Gilbreth2021,Gilbreth2015}.  Our low-temperature results for the condensate fraction show remarkable agreement with the $T=0$ QMC result of Ref.~\cite{Astrakharchik2005}. Using our finite-temperature results for the condensate fraction for multiple particle numbers $N$, we used finite-size scaling to determine the superfluid critical temperature. Our result $T_c=0.16(1)\;T_F$ is in agreement with experiment~\cite{Ku2012} and with the QMC results of Refs.~\cite{Burovski2006, Goulko2010}. 
 
To address open questions regarding the existence of a pseudogap regime and its extent in the UFG, we calculated a model-independent energy-staggering pairing gap and the spin susceptibility in the continuum limit. The calculation of the energy-staggering pairing gap was made possible by the use of a canonical-ensemble AFMC formulation of fixed particle numbers.  The pairing gap and spin susceptibility can be used to identify signatures of a pseudogap regime above $T_c$ and below the pairing temperature scale $T^*$.  From the largest particle number $N=114$ simulations in the continuum limit we find an upper bound of $T^*\approx 0.2\,T_F$. Thus we conclude that the pseudogap regime in the UFG is narrow.  Compared with our previous finite filling factor study in Ref.~\cite{Jensen2020a}, we observe  that in the continuum limit the pairing signatures increased for fixed temperature and particle number $N$. This is consistent with the results of Ref.~\cite{Gezerlis2008} where it was shown at $T=0$ that a finite effective range interaction suppresses attractive pairing correlations. 

We also calculated the free energy staggering pairing gap and found it to have significantly reduced statistical errors compared with the energy-staggering pairing gap. Our results for this observable supports a pairing temperature scale of  $T^{*}\approx 0.2\;T_F$.  In addition, extrapolating the low-temperature results for the free energy gap to zero temperature, we obtained an accurate estimate for the zero-temperature pairing gap of $\Delta_E = 0.576(24) \, \varepsilon_F$. Such extrapolation is not possible with the energy-staggering pairing gap because of its much larger statistical errors.

 While we have reached the continuum limit for the first time for several thermodynamic observables that are useful signatures of pairing correlations for particle number as large as $N=114$, we have not quite reached the thermodynamic limit of large particle number for the pairing gap and the spin susceptibility. It would be interesting to investigate the thermodynamic limit of these observables in future works. 

\acknowledgments
This work of was supported in part by the U. S. DOE grants Nos.~DE-SC0019521, DE-SC0020177, and DE-FG02-00ER41132.
The calculations presented here used resources of the National Energy Research Scientific Computing Center (NERSC), a U.S. Department of Energy Office of Science User Facility operated under Contract No.~DE-AC02-05CH11231.  We also thank the Yale Center for Research Computing for guidance and use of the research computing infrastructure.

\appendix

\section{Continuous time and continuum extrapolations}~\label{app:extrapolations}

We demonstrate the continuous time and continuum extrapolations for the condensate fraction in Sec.~\ref{ext-n}. Typical continuum extrapolations for the free energy gap are shown in Sec.~\ref{ext-Fgap}.\\

\begin{figure*}[h]
\includegraphics[scale=0.84]{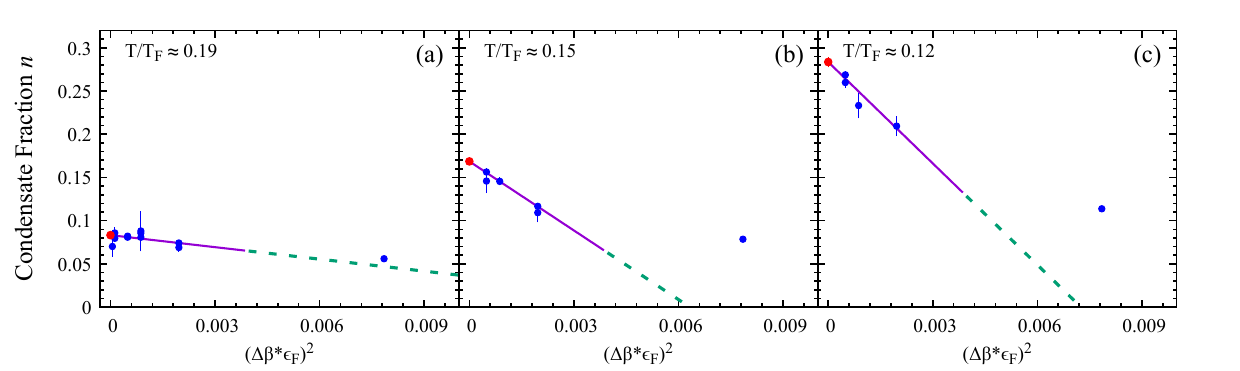}
\caption{Extrapolation in $\Delta \beta$ for $N=40$ particles and lattice size $N_{L}^3=9^3$ or the condensate fraction $n$ at temperatures (a) $T/T_{F}\approx 0.19$, (b) $T/T_{F}\approx 0.15$, and (c) $T/T_{F}\approx 0.12$.  The solid blue circles are the calculated condensate fraction for several discrete values of $\Delta\beta$, We perform a straight line fit (solid purple line) in $(\Delta \beta*\varepsilon_{F})^2$ (where $\varepsilon_{F}=(\hbar^2/2m)(3\pi^2\rho)^{2/3})$ is the Fermi energy) in the range  $(\Delta \beta*\varepsilon_{F})^2 < 0.003$. The fitted line is shown to larger values of $\Delta\beta$ for reference (dashed green line).  
}\label{fig:dbeta}
\end{figure*}

\begin{figure*}[h]
\includegraphics[scale=0.492]{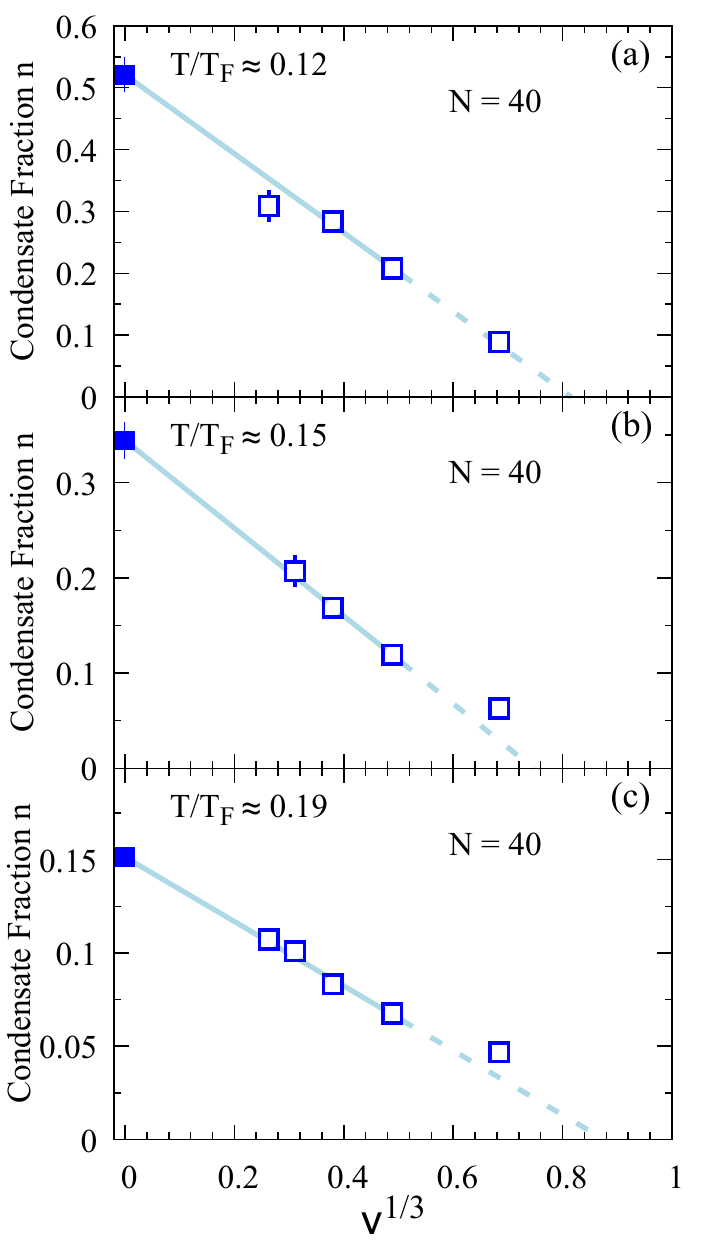}
\includegraphics[scale=0.492]{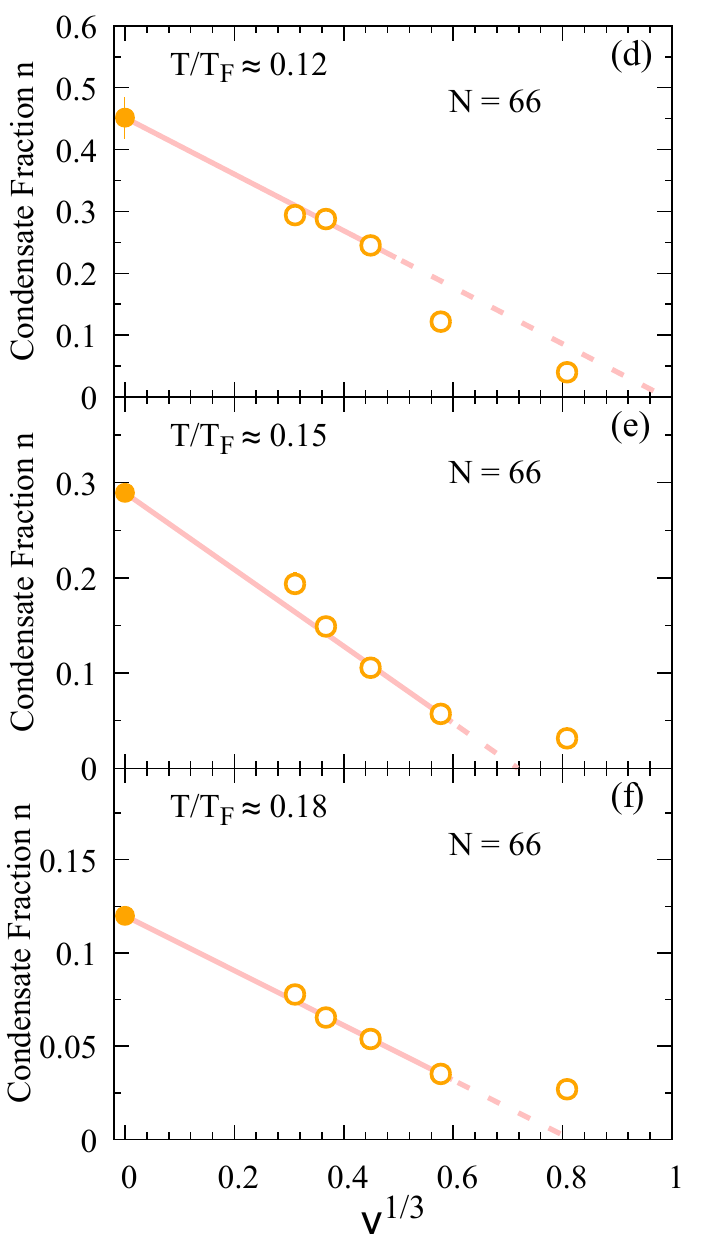}
\includegraphics[scale=0.492]{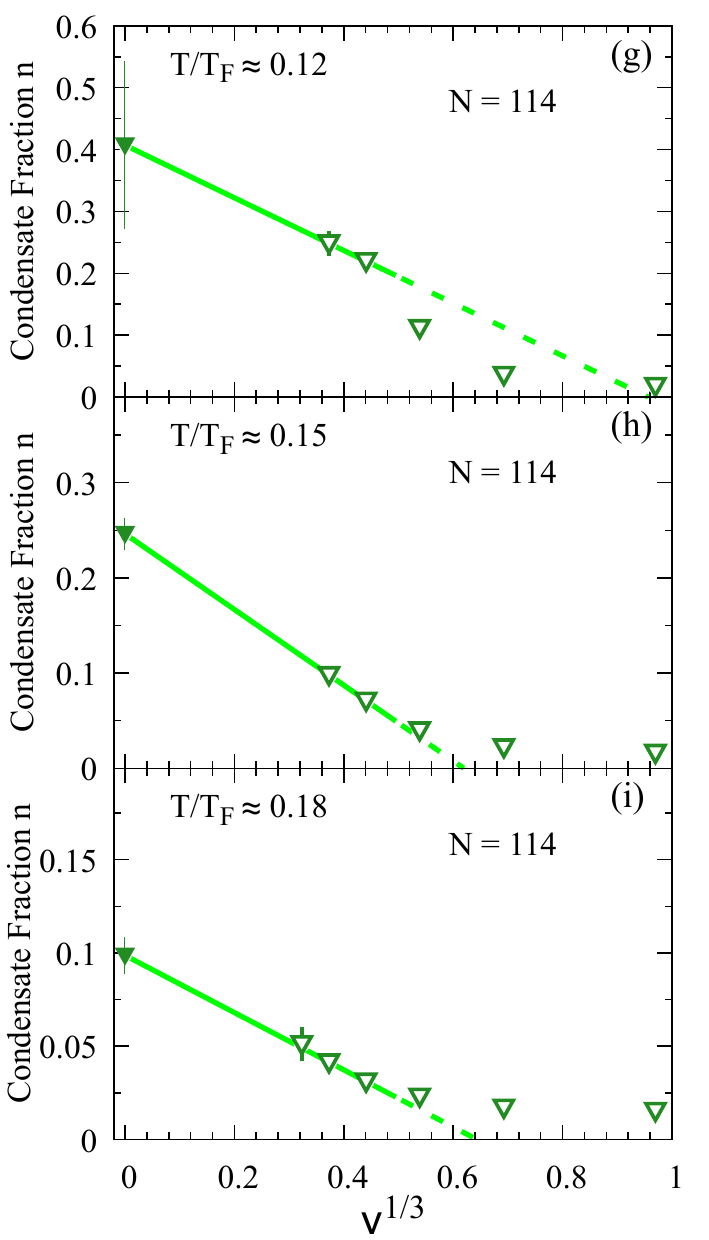}
\caption{Continuum limit extrapolations $\nu \rightarrow 0$ ($\nu=N_L^3/N$ is the filling factor) of the condensate fraction for $N=40$ particles (left column, open blue squares), $N=66$ particles (middle column, open orange circles), and $N=114$ particles (right column, open down green triangles) for several representative temperatures $T/T_{F}\approx 0.12$ (top row),  and $T/T_{F}\approx 0.15$ (middle row). In the bottom row we show the extrapolations for $T/T_{F}\approx 0.19$ ($N=40$) and $T/T_{F}\approx 0.18$ ($N=66, 114$).  We perform a straight line fit  (solid line) in $\nu^{1/3}$ in the range $\nu^{1/3} < 0.5$ for each temperature $T/T_{F}$ and particle number $N$. The fitted line is shown to larger filling factors $\nu^{1/3}$ for reference (dashed line).}\label{fig:dens_extrap_ODLRO} 
\end{figure*}

\begin{figure*}[bth]
\includegraphics[scale=0.44]{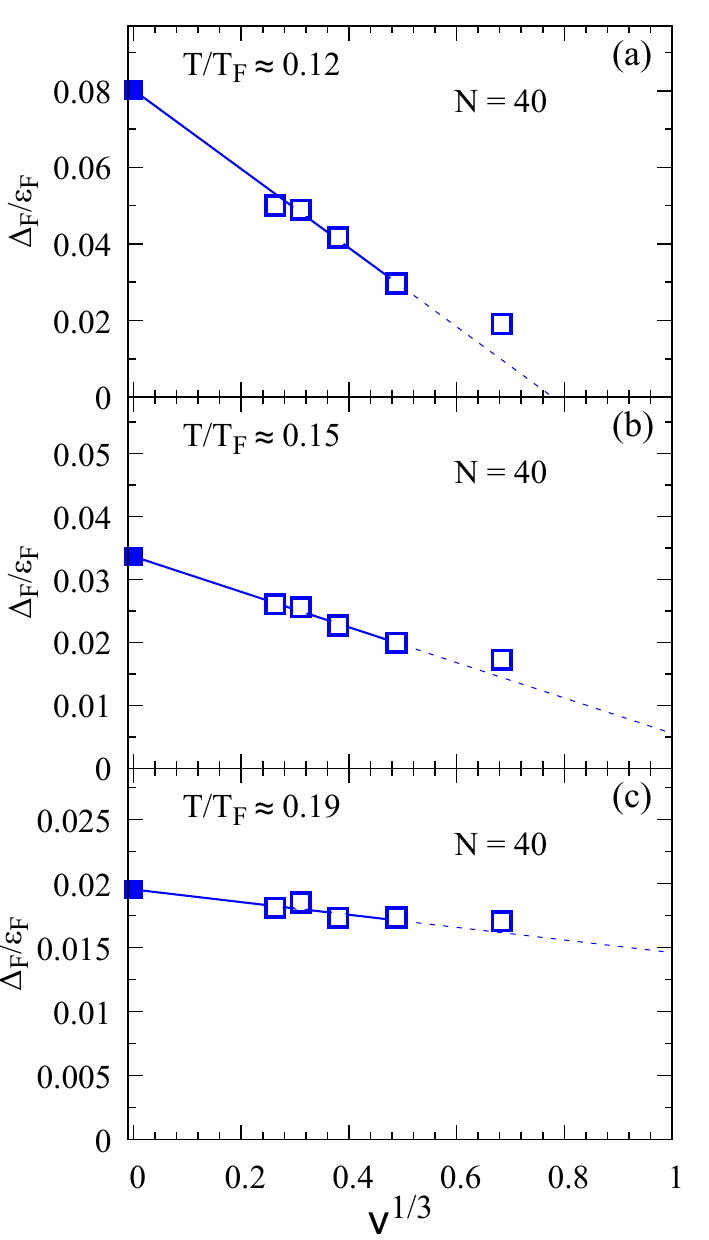}
\includegraphics[scale=0.44]{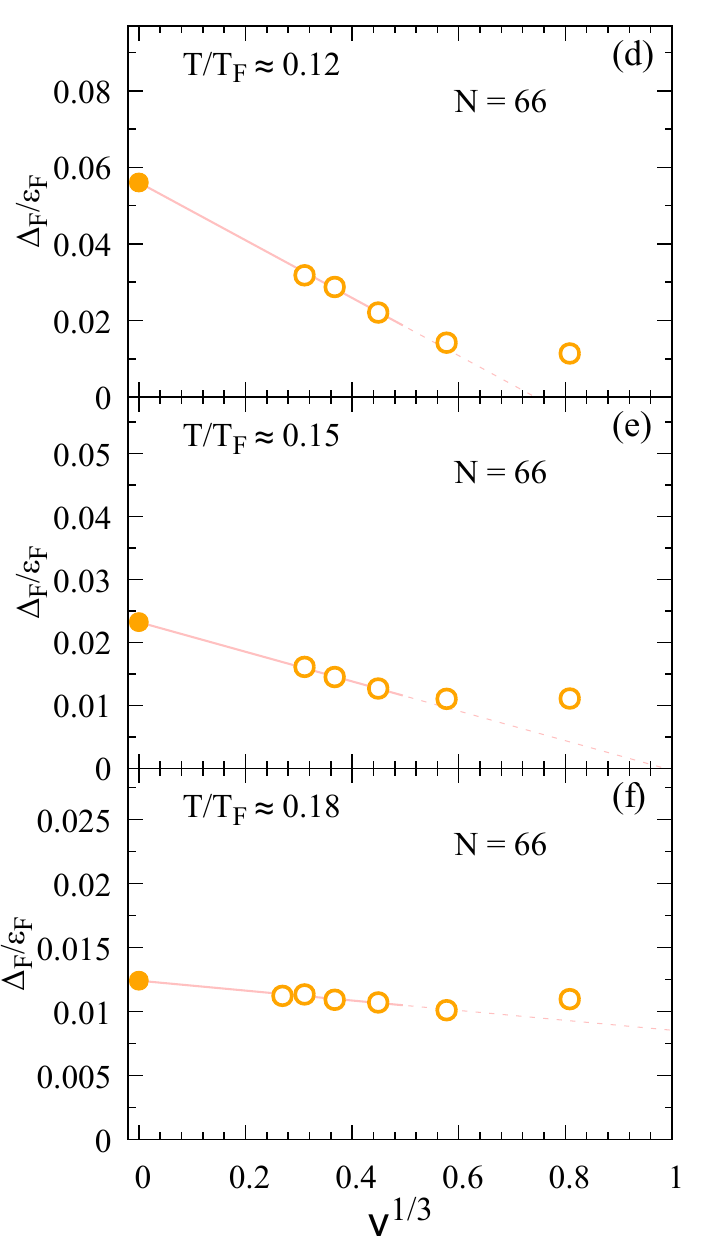}
\includegraphics[scale=0.44]{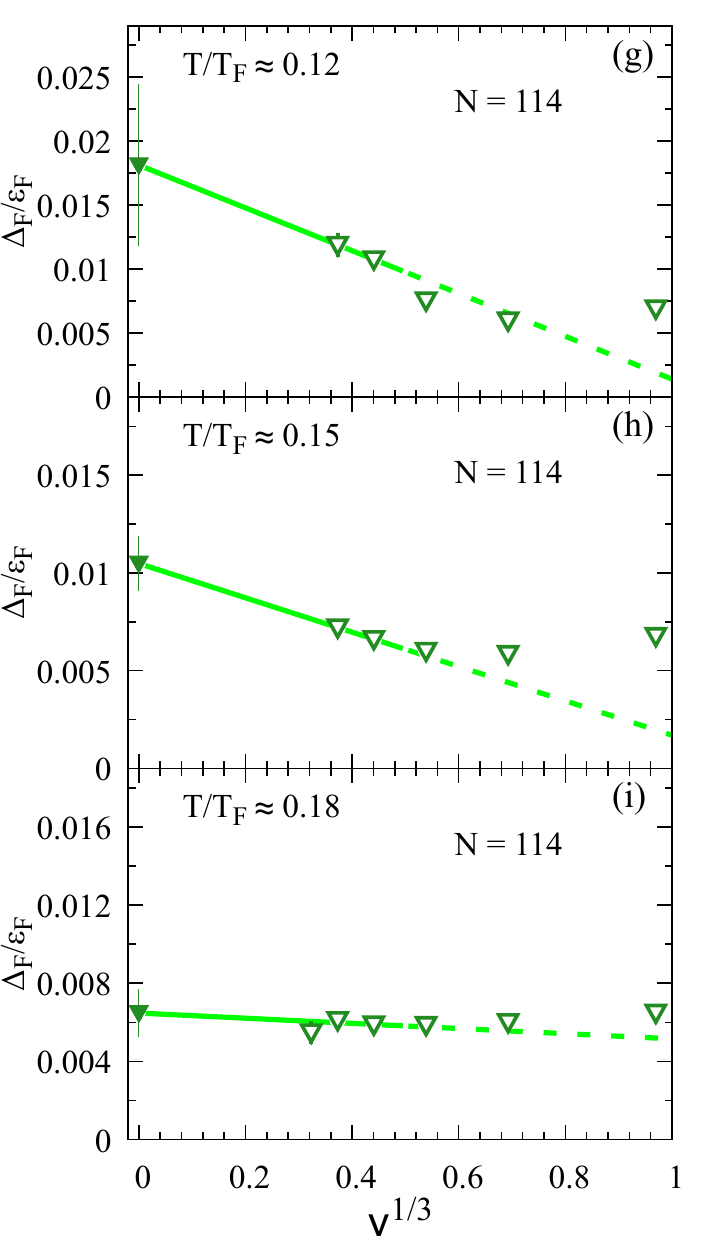}
\caption{Continuum limit extrapolations in $\nu^{1/3}$ for the free energy-staggering pairing gap $\Delta_F$ for $N=40, 66$ and $114$ particles and for a few representative temperatures as in Fig.~\ref{fig:dens_extrap_ODLRO}. We perform straight line fits (solid lines) in the range $\nu^{1/3} < 0.5$.}\label{fig:dens_extrap_fgap}
\end{figure*}

\subsection{Condensate fraction} \label{ext-n}

\ssec{Extrapolations to continuous time} 
The finite imaginary time slice $\Delta \beta$ introduces a systematic error in the AFMC calculations. As discussed in Sec.~\ref{HS_trans}, using a symmetric Trotter decomposition and a three-point Gaussian quadrature leads to a discretization error of $O(\Delta\beta^{2})$.  We extrapolate $\Delta\beta\rightarrow 0$ by a linear fit in $(\varepsilon_F \Delta \beta)^2$ for sufficiently small values of $\Delta\beta$. 

In Fig.~\ref{fig:dbeta} we show representative extrapolations of the condensate fraction for $N=40$ particles and lattice size of $9^3$ at three  temperatures (a) $T/T_{F}\approx 0.19$, (b) $T/T_{F}\approx 0.15$, and (c) $T/T_{F}\approx 0.12$. The slope of the fit increases dramatically in magnitude as temperature is reduced, so the extrapolations become necessary at low temperatures (see for $T/T_{F}\approx 0.12$ and  $T/T_{F}\approx 0.15$). We performed these $\Delta\beta$ extrapolations for all particle numbers and lattice sizes. \\

\ssec{Continuum extrapolations} 
After carrying out the $\Delta\beta$ extrapolations, we perform the continuum limit extrapolation using lattices of increasing size with $\nu \to 0$ for each temperature $T/T_{F}$ and particle number $N$.  

In Fig.~\ref{fig:dens_extrap_ODLRO} we show the condensate fraction $n$ as a function of $\nu^{1/3}$ for particle numbers $N=40, 66$ and $114$ for three temperatures in the critical regime near $T_c$. We perform linear fits in $\nu^{1/3}$ using the range $\nu^{1/3}\lesssim 0.5-0.6$.  We use this range in the continuum extrapolations of the condensate fraction $n$ for all temperatures and particle numbers. We note that the behavior of $n$ for larger filling factors is significantly different for lower temperatures.  At the lower temperature $T/T_{F}\approx 0.12$  (upper panels of Fig.~\ref{fig:dens_extrap_ODLRO}), the largest filling factor results are below the continued straight line fit (dashed line) whereas for higher temperatures (middle and lower panels), the larger filling factor results are above the continued fit. The continuum extrapolation provides the final estimate for the condensate fraction $n$ for a given particle number $N$ and temperature $T/T_F$. 

\subsection{Free energy gap}\label{ext-Fgap}

In this section, we demonstrate the continuum limit ($\nu \rightarrow 0$) extrapolations for the free energy gap $\Delta_F$.  These continuum extrapolations are carried out after the continuous time extrapolations $\Delta \beta \rightarrow 0$. 

 In Fig.~\ref{fig:dens_extrap_fgap}  we show $\Delta_F$ vs.~$\nu^{1/3}$ for several values of particle number and temperatures (as in Fig.~\ref{fig:dens_extrap_ODLRO}).   The straight lines are fits in the range $\nu^{1/3} < 0.5$.

\newpage

%


\end{document}